\documentclass[11pt]{article}
\usepackage{amssymb,latexsym,amsmath,amsthm,graphicx,subfigure}
\usepackage{bm}
\def\la {\lambda}

\makeatletter
\@addtoreset{figure}{section}
\def\thefigure{\thesection.\@arabic\c@figure}
\def\fps@figure{h, t}
\@addtoreset{table}{bsection}
\def\thetable{\thesection.\@arabic\c@table}
\def\fps@table{h, t}
\@addtoreset{equation}{section}

\newif\ifamsfonts\amsfontstrue \ifamsfonts
\font\twlbbb=msbm10 scaled\magstep1 \font\egtbbb=msbm8
\font\sixbbb=msbm6
\newfam\mathbbfam
\textfont\mathbbfam=\twlbbb \scriptfont\mathbbfam=\egtbbb
\scriptscriptfont\mathbbfam=\sixbbb
\makeatother

\newtheorem{theorem}{Theorem}[section]

\newtheorem{lemma}[theorem]{Lemma}

\newfont{\tenbi}{cmbxti10}

\begin{document}

\title{Algebraic Closed Geodesics on a Triaxial Ellipsoid
\footnote{AMS Subject Classification 14H52, 37J45, 53C22, 58E10}}
\author{Yuri N. Fedorov \\
 Department of Mathematics and Mechanics
 \\ Moscow Lomonosov University, Moscow, 119 899, Russia \\
e-mail: fedorov@mech.math.msu.su \\
and \\
 Department de Matem\`atica I, \\
Universitat Politecnica de Catalunya, \\
Barcelona, E-08028 Spain \\
e-mail: Yuri.Fedorov@upc.edu}
\maketitle

\abstract{We propose a simple method of explicit description of families of closed geodesics on
a triaxial ellipsoid $Q$ that are cut out by algebraic surfaces in ${\mathbb R}^3$.
Such geodesics are either connected components of spatial elliptic curves or of rational
curves.

Our approach is based on elements of the Weierstrass--Poncar\'e reduction theory for
hyperelliptic tangential
covers of elliptic curves, the addition law for elliptic functions, and
the Moser--Trubowitz isomorphism between geodesics on a quadric and stationary solutions of the
KdV equation.
For the case of 3-fold and 4-fold coverings, explicit formulas for the cutting
algebraic surfaces are provided and some properties of the corresponding geodesics
are discussed. }

\section{Introduction}
One of the best known classical integrable systems is
the geodesic motion on a triaxial ellipsoid $Q\subset {\mathbb R}^3$.
By introducing ellipsoidal coordinates on $Q$, the problem was
reduced to hyperelliptic quadratures by Jacobi (\cite{Jac})
and was integrated in terms of theta-functions of a genus 2 hyperelliptic
curve $\Gamma$ by Weierstrass in \cite{Weier}.

A generic geodesic on $Q$ is known to be quasiperiodic and
oscillate between 2 symmetric curvature lines (caustics).

It is of a certain interest to find conditions for a geodesic on
$Q$ to be periodic (closed) and to describe such geodesics
explicitly. At first sight this problem has a standard solution: by
introducing action--angle variables $\{I_1, I_2, \phi_1, \phi_2\}$ one
can define frequencies
$$
\dot\phi_1=\Omega_1(I_1, I_2), \quad  \dot\phi_2=\Omega_2(I_1, I_2).
$$
Then the geodesic is closed if and only if the rotation number $\Omega_2/\Omega_1$
is rational.

However, the frequencies $\Omega_j$ are known to be linear
combinations of Abelian integrals on the hyperelliptic curve,
hence the condition on the rotation number implies a
transcendental equation on the constants of motion and the
parameters of the problem. In practice, this appears to be useless
for an exact description of closed geodesics.

Such a description can be made much more explicit when the
hyperelliptic curve $\Gamma$ turns out to be a covering of an
elliptic curve $\cal E$ and a certain holomorphic differential
reduces to a holomorphic differential on $\cal E$. Then the
corresponding geodesic itself is a spatial elliptic curve, which
covers $\cal E$\footnote{More precisely, it is a connected
component of a real part of an elliptic curve}, or a rational
curve and, as an algebraic subvariety in ${\mathbb R}^3$
(${\mathbb C}^3$), it can be represented as a connected component
of the intersection of the ellipsoid $Q$ with an algebraic
surface. In the sequel, such class of geodesics will be referred
to as {\it algebraic closed geodesics}.

Conversely, one can also show that any algebraic closed geodesic on an ellipsoid
must be a connected component of an elliptic or a rational curve.

Studying closed geodesics on quadrics is a classical problem.
Surprisingly, we did not find any reference to its explicit
solution in the classical or modern literature. Here we can only
quote the paper \cite{Braun}, which studied the case of a 2-fold
covering of an elliptic curve, when the solution is expressed in
terms of two elliptic functions of time with different period
lattices. Since the periods are generally incommensurable, the corresponding
geodesics are not periodic but quasi-periodic.

Note that in the problem of periodic orbits of the Birkhoff billiard inside an ellipsoid
much more progress has been made (see \cite{Drag_Rad_main, Drag_Rad, Emma0}).

\paragraph{Contents of the paper.}
The paper proposes a simple approach to explicit description of algebraic surfaces $\cal V$
in ${\mathbb R}^3$ that cut out closed geodesics on $Q$. It is based on  elements of the
Weierstrass--Poncar\'e theory of reduction of Abelian functions
(see, e.g.,\cite{Krazer,Bel,Enol}), addition law for elliptic functions, and
the remarkable relation between geodesics on a quadric and stationary solutions of the
KdV equation (the Moser--Trubowitz isomorphism) described in \cite{Knorr2} and
recently revisited in \cite{AF} in connection with periodic orbits of geodesic billiards
on an ellipsoid.

Namely, for each genus 2 hyperelliptic tangential cover of an elliptic curve we construct
a one-parameter family of plane algebraic curves (so called {\it polhodes}).
Appropriate connected components of the polhodes have a form of Lissajou curves and
describe closed geodesics in terms of the two ellipsoidal
coordinates $\lambda_1, \lambda_2$ on $Q$.
The geodesics of one and the same family are tangent to the same caustic on $Q$ and
the parameters of the ellipsoid are functions of the moduli of the elliptic curve.

Since equations of the polhodes 
depend only on the symmetric functions $\lambda_1+\lambda_2$,
$\lambda_1\lambda_2$, they can be rewritten in terms of Cartesian
coordinates in ${\mathbb R}^3$ thus giving the above mentioned
algebraic surfaces $\cal V$.

Our family of closed geodesic contains special ones, which have mirror symmetry with respect to principal
coordinate planes in ${\mathbb R}^3$. In particular, for the case of 3-fold and 4-fold coverings of
an elliptic curve, the special geodesics are cut out by quadratic and, respectively,
cubic surfaces in ${\mathbb R}^3$, as illustrated in Figures \ref{sp1.fig} and \ref{special4_1.fig}.

Depending on how to assign the parameters of the ellipsoid and of the
caustic to the branch points of the hyperelliptic curve, one can
obtain closed geodesics with or without self-intersections on $Q$.

\section{Linearization of the geodesic flow on the \\ ellipsoid and some
elliptic solutions}

We first briefly recall the integration of the geodesic motion
on an $n$-dimensional ellipsoid
\begin{gather*}
Q = \left\{ \frac{X_1^2}{a_1}+\cdots +\frac{X_{n+1}^2}{a_{n+1}}= 1
\right\}\subset {\mathbb R}^{n+1}=(X_1,X_2,\dots,X_{n+1}), \\
0 < a_1 < \cdots < a_n< a_{ n + 1} .
\end{gather*}
Let $t$ be the natural
parameter of the geodesic and $\la_1,\dots,\la_n$ be the
ellipsoidal coordinates on $Q$ defined by the formulas
\begin{equation}
\label{spheroconic2}
X_i^2 = a_i \frac{(a_i-{\la}_1)\cdots (a_i-{\la}_n)}
{ \prod_{j\ne i}(a_i-a_j)} \, , \qquad i=1,\dots, n+1 .
\end{equation}
In these coordinates and their derivatives $d\lambda_k/d t$ the
total energy $\displaystyle \frac 12 (\dot X, \dot X)$ takes a
St\"ackel form, and after time re-parameterization \footnote{For
$n=2$ this re-parameterization was made by Weierstrass
\cite{Weier}.}
\begin{equation} \label{tau-1}
dt =\la_1\cdots\la_n\, ds ,
\end{equation}
the evolution of $\lambda_{i}$ is described by quadratures
\begin{gather}
\frac{\la_1^{k-1} d\la_1}{2\sqrt{- R(\la_1)}} +\cdots
+ \frac{\la_n^{k-1} d\la_n}{2\sqrt{-R (\la_n)}}
 =\Bigg \{
\begin{aligned}
ds\quad \mbox{ for } & k=1\, , \\
0\quad \mbox{ for } & k=2, \dots, n ,
\end{aligned} \label{quad2} \\
{R}(\la) =\la (\la-a_1)\cdots (\la-a_{n+1}) (\la-c_1)\cdots (\la-c_{n}), \nonumber
\end{gather}
where $c_k$ are constants of motion.

This implies integrability of the system by the Liouville theorem.
The generic invariant varieties of the flow are $n$-dimensional
tori with a quasiperiodic motion. The corresponding geodesics are
tangent to one and the same set of $n-1$ confocal quadrics
$Q_{c_1},\dots,Q_{c_{n-1}}$ of the confocal family
\begin{equation} \label{family}
Q_c = \left\{ \frac{X_1^2}{a_1-c}+\cdots +\frac{X_{n+1}^2}{a_{n+1}-c}= 1
\right\}\subset {\mathbb R}^{n+1}.
\end{equation}
In particular, generic geodesics on a 2-dimensional ellipsoid
fill a ring bounded by caustics,
the lines of intersection of $Q$ with confocal hyperboloid $Q_{c_1}$.

The quadratures (\ref{quad2}) involve $n$ independent holomorphic differentials
on the genus $n$ hyperelliptic curve
$\Gamma=\{\mu^2=- R(\la)\}$,
\begin{equation} \label{omegas}
\omega_k =\frac{\la^{k-1} d\la }{2 \sqrt{ {R} (\la) }}\, , \qquad k=1,\dots, n
\end{equation}
and give rise to the
Abel--Jacobi map of the $n$-th symmetric product $\Gamma^{(n)}$ to the Jacobian variety 
of $\Gamma$,
\begin{gather}
\int \limits^{P_1}_{P_0} \omega_k + \cdots + \int
\limits^{P_n}_{P_0} \omega_k =u_{k}, \qquad
P_k=\left(\lambda_k,\sqrt{-{R} (\la_k)} \right ) \in \Gamma,  \label{AB}
\end{gather}
where $u_1,\dots,u_n$ are coordinates on the universal covering of
Jac$(\Gamma)$ and $P_0$ is a fixed basepoint, which we choose to
be the infinity point $\infty$ on $\Gamma$.

Since $u_1=s +$const and $u_2,\dots,u_n=$const, the geodesic
motion in the new parameterization is linearized on the Jacobian
variety of $\Gamma$.

The inversion of the map (\ref{AB}) applied to formulas
(\ref{spheroconic2}) leads to the following parameterization of a
generic geodesic in terms of $n$-dimensional theta-functions
$\theta(w_1,\dots,w_n)$ associated to the curve $\Gamma$,
\begin{gather} \label{theta_n}
X_i (s ) = \varkappa_i \, \frac{ \theta[\eta_i] (w_1,\dots,w_n) }
{\theta[\Delta] (w_1,\dots,w_n) }, \qquad i=1,\dots, n+1,
\end{gather}
where $\Delta=(\delta'',\delta')$,
$\eta_{i}=(\eta_{i}'',\eta_{i}')\in {\mathbb R}^{2n}/2 {\mathbb
R}^{2n}$ are certain half-integer theta-characteristics, the arguments
$w_1,\dots,w_n$ depend linearly on $u_1,\dots,u_n$, and therefore on $s$,
and $\varkappa_i$ are constant factors depending on the moduli of $\Gamma$ only.

For the classical Jacobi problem $(n=2)$, the complete
theta-functional solution was presented in \cite{Weier}, and, for
arbitrary dimensions, in \cite{Knorr}, whereas a complete
classification of real geodesics on $Q$ was made in \cite{Audin}.

\paragraph{Periodicity problem and a solution in terms of elliptic functions.}
As mentioned in Introduction, we restrict ourselves with the case when
a geodesic is periodic in the complex parameter $s$, namely, double-periodic.
This implies that the solution (\ref{theta_n}) can be expressed in terms of elliptic functions
of $s$.

As an example, following von Braunmuhl \cite{Braun}, consider the
geodesic problem on 2-dimensional quadric ($n=2$) and suppose that
the parameters $a_i, c_j$ in (\ref{quad2}) are such that the curve
$\Gamma$ becomes birationally equivalent to the following
canonical curve
$$
w^2 = - z(z-1)(z-\alpha)(z-\beta)(z-\alpha \beta),
$$
$\alpha, \beta$ being arbitrary positive constant.
Then, as widely described in the literature (see, e.g., \cite{Krazer,Bel,Enol}),
$\Gamma$ covers two different elliptic curves
$$
\ell_\pm = \{W_\pm^2=Z_\pm (1-Z_\pm)(1-k_\pm Z_\pm) \}, \qquad
k_\pm^2=- \frac {(\sqrt{\alpha}\mp \sqrt{\beta} )^2}{(1-\alpha)(1-\beta)}
$$
with covering relations
\begin{align}
Z_+=Z_- & = \frac{ (1-\alpha)(1-\beta)z}{ (z-\alpha)(z-\beta) }, \label{2-fold}\\
W_\pm & = -\sqrt{(1-\alpha)(1-\beta)}\,
\frac {z\mp \sqrt{\alpha\beta}}{(z-\alpha)^2(z-\beta)^2 }\, w.\nonumber
\end{align}
Thus, $\Gamma$ is a 2-fold covering of $\ell_-$ and $\ell_+$.

Both holomorphic differentials $\omega_1, \omega_2$ on
$\Gamma$ reduce to linear combinations of the holomorphic
differentials on $\ell_+$ and $\ell_-$, namely
$$
\frac{d Z_\pm}{W_\pm} = \frac{z\mp \sqrt{\alpha \beta}}{w}\, dz .
$$
Then a linear combination of equations (\ref{quad2}) for $n=2$ yields
\begin{align*}
\int_{\infty}^{\la_1} \frac{z- \sqrt{\alpha \beta} }{w}\,dz +
\int_{\infty}^{\la_2} \frac{z- \sqrt{\alpha \beta}}{w}\, dz & =- s\sqrt{\alpha \beta}
 +\mbox{const}\, , \\
\int_{\infty}^{\la_1} \frac{z+ \sqrt{\alpha \beta}}{w}\,dz +
\int_{\infty}^{\la_2} \frac{z+ \sqrt{\alpha \beta}}{w}\,dz & =\; s\sqrt{\alpha \beta}
 +\mbox{const}\, .
\end{align*}
Inversion of these quadratures lead to solutions for $X_i$ in terms of elliptic
functions of the curves $\ell_\pm$, whose arguments both depend on the time parameter $s$.
Then, since their periods are generally
incommensurable, the corresponding geodesics remain to be quasi-periodic.

This observation shows that not any case of covering $\Gamma$ to an elliptic curve
results in closed geodesics on $Q$. In the next section we consider other types of coverings
and obtain sufficient condition for a geodesic to be an elliptic curve.

\section{Hyperelliptic tangential covers and closed \\
geodesics on an $n$-dimensional ellipsoid}

Consider a genus $n$ compact smooth hyperelliptic surface $G$, whose affine
part $G_A\subset {\mathbb C}^2=(z,w)$ is given by equation
$$
w^{2}= - R_{2n+1}(z) ,$$ $R_{2n+1}(z)$ being a polynomial of
degree $2n+1$. The curve $G$ is obtained from $G_A$ by gluing the
infinite point $\infty$. Let $\{ \varOmega_1(P), \dots,
\varOmega_n (P)\}$, $P=(z,w)\in G$ be a basis of independent
holomorphic differentials on $G$. One can also write
$\varOmega_j(P)=\phi_j(P)\, d\tau$, where $\tau$ is a local
coordinate in a neibourhood of $P$. Next, let $\Lambda$ be the
lattice in ${\mathbb C}^{n}$ generated by $2n$ independent period
vectors $\oint (\varOmega_1, \dots, \varOmega_n )^T$.

The curve $G$ admits a canonical embedding into its Jacobian variety
Jac$(G)={\mathbb C}^{n} (u_1,\dots,u_n)/\Lambda$,
\begin{equation} \label{embed}
P\mapsto {\cal A}(P)= \int_{\infty}^{P} (\varOmega_1, \dots, \varOmega_n )^T,
\end{equation}
so that $\infty$ is mapped into the neutral point (origin) in Jac$(G)$ and
$$
{\bf U}=\frac d{d\tau}{\cal A(P)} \bigg |_{P=\infty}
=(\phi_1(\infty), \dots, \phi_n(\infty))^T
$$
is the tangent vector of $G\subset$Jac$(G)$ at the origin.

Now assume that $G$ is an $N$-fold covering of an elliptic curve ${\cal E}$, which we represent in
the canonical Weierstrass form
\begin{equation} \label{ell1}
{\cal E}= \left\{(\wp' (u))^2
= 4\wp^{3}(u)-g_{2}\wp(u)-g_{3}
\equiv 4 (\wp-e_1 )(\wp-e_2)(\wp-e_3) \right \} .
\end{equation}
Here $\wp(u)=\wp (u\mid \omega,\omega')$ denotes the Weierstrass elliptic function with
half-periods $\omega,\omega'$ and $u\in {\mathbb C}/\{ {\mathbb Z}\omega+{\mathbb Z}\omega'\}$.
The parameters $g_2,g_3$ provide moduli of the curve.

Assume also that under the covering map $\pi \, :\, G\mapsto {\cal
E}$ the infinite point $\infty\in G$ is mapped to $u=0$.

In the sequel we concentrate on {\it hyperelliptic tangential coverings}
$\pi \, :\, G\mapsto {\cal E}$, when
${\cal E}$ admits the following canonical embedding onto Jac$(G)$
$$
u\mapsto (u_1,\dots,u_n)=u {\bf U} .
$$
That is, the images of ${\cal E}$ and $G$ in Jac$(G)$ are tangent at the origin
\footnote{As follows from this definition,
the 2-fold covers (\ref{2-fold}) are not hyperellipticallly tangential.}.
The motion of tangential covering was introduced in \cite{TV, Tr} in connection
with elliptic solutions of the KdV equation (see also \cite{Bel, Enol, Smirnov2}).

Namely, let $\theta(u_1,\dots,u_n)$ be the theta-function associated to the covering curve $G$
and $\Theta$ be the theta-divisor, codimension one subvariety of ${\rm Jac} (\Gamma)$ defined
by equation $\theta[\Delta]({\bf u})=0$,
where $[\Delta]$ is the special theta-characteristic in the solution (\ref{theta_n}).

\begin{theorem} \label{N-section}
\textup{(\cite{Krich})}
For an arbitrary vector ${\bf W}\in {\mathbb C}^g$, the transcendental equation
\begin{equation} \label{ECM}
\theta ({\bf U}x+ {\bf W} )=0, \qquad x \in {\mathbb C},
\end{equation}
has exactly $N$ solutions $x=q_1({\bf W}), \dots, x=q_N({\bf W})$ (possibly, with multiplicity).
\end{theorem}


That is, the complex flow on Jac$(G)$ in ${\bf U}$-direction
intersects the theta-divisor $\Theta$ or any its translate
at a finite number of points. This property is exceptional:
for a generic hyperelliptic curve $G$ the number of such intersections is infinite.

Note that in the local coordinates $u_1,\dots, u_n$ on Jac($G$) corresponding
to the standard basis of holomorphic differentials
\begin{equation} \label{new-omegas}
{\bar\omega}_k =\frac{z^{k-1}\, dz }{w}\, , \qquad k=1,\dots, n,
\end{equation}
one has ${\bf U}=(0,\dots,0,2)^T$.

According to the Poincar\'e reducibility theorem (see e.g.,
\cite{Bel}), apart from the curve $\cal E$, the Jacobian of
$G$ contains an $(n-1)$-dimensional Abelian subvariety ${\cal A}_{n-1}$.
For $n=2$ the subvariety is just another elliptic curve covered by $G$.

Notice that for the case $n=2$, explicit algebraic expressions of the
covers and coefficients of hyperelliptic curves are known for
$N\le 8$ (see \cite{Tr}).

\paragraph{Double periodic geodesics on an ellipsoid.}
The algebraic geometrical property described by Theorem \ref{N-section} gives a tool for
a description of double-periodic geodesic flow on the $n$-dimensional quadric $Q$, which is
linearized on the Jacobian of the hyperelliptic curve $\Gamma$ in Section 2.
Namely, let the genus $n$ curves $G$ and $\Gamma$
are related via birational transformation of the form
\begin{equation} \label{la-z}
\lambda=\frac{\alpha}{(z-\beta)}, \quad \mu= \frac w{(z-\beta)^{n+1}},
\end{equation}
where $(\beta,0)$ is a finite Weierstrass point on $G$ and $\alpha$ is an arbitrary positive
constant.
Then the following theorem proved in \cite{AF} holds.

\begin{theorem} \label{KdV-Jacobi}
To any hyperelliptic tangential cover $G\mapsto {\cal E}$ such that
all the Weierstrass points of $G$ are real, one can associate an $(n-1)$-parametric
family of different closed real geodesics on an $n$-dimensional ellipsoid $Q$ that are tangent
to the same set of confocal quadrics $Q_{c_1},\dots,Q_{c_{n-1}}$. The parameters of
the ellipsoid ($a_i$) and of the quadrics ($c_j$)
are related to branch points of $G$ via the transformation (\ref{la-z}).
\end{theorem}

\paragraph{Remark.} It is natural to consider a closed geodesic as a curve on $Q$ and
not as a periodic solution of the geodesic equations that depends on the initial point
on the curve as on a parameter. That is, we disregard this parameter in the above family
of closed real geodesics.
\medskip

\noindent{\it Proof of Theorem} \ref{KdV-Jacobi}. The transformation
(\ref{la-z}) sends the points $\infty$ and $(\beta,0)$ on $G$ to the Weierstrass points
${\cal O}=(0,0)$ and, respectively, $\infty$ on $\Gamma$.
Then, identifying the
curves $G$ and $\Gamma$, as well as their Jacobians, we find
that the ${\bf U}$-flow on Jac($G$), which is tangent to the canonically
embedded hyperelliptic curve $\Gamma\subset \mbox {Jac}({\Gamma})$
at $\infty$, is represented as the flow on Jac($\Gamma$) which is tangent to
the embedded $\Gamma\subset{\rm Jac}(\Gamma)$ at ${\cal O}$, and vice versa.
In the coordinates on  Jac$(\Gamma)$  corresponding to the basis (\ref{omegas}),
the latter flow has direction $(1,0,\dots,0)^T$ and thus coincides with the linearized
geodesic flow on $Q$.

This remarkable relation was first described in \cite{Knorr2} as
the Moser--Trubowitz isomorphism between stationary $n$-gap solutions of the
KdV equation and generic (quasiperiodic) geodesics on an $n$-dimensional quadric.

Next, let us fix a real constant $d$ and the confocal quadric $Q_d$ of the family
(\ref{family}) such that the geodesics with the constants of motion
$c_1,\dots,c_{n-1}$ have a non-empty intersection with $Q\cap Q_d$.
In view of (\ref{spheroconic2}),
when a geodesic $X(s)$ intersects $Q\cap Q_d$, one of the points
$P_i=(\lambda_i,\mu_i)$ on the curve $\Gamma$ (without loss of generality we
choose it to be $P_n$) coincides with one of the points $E_{d
\pm}=(d,\pm \sqrt{R(d)})$. Under the Abel--Jacobi map (\ref{AB}) with $P_0=\infty$,
the condition $P_n=E_{d\pm}$  defines two translates of the theta-divisor
$$
\Theta_{d\pm} =\{ \theta[\Delta]({\bf u} \mp q/2)=0 \} \subset {\rm Jac}(\Gamma), \qquad
 q = \int_{E_{d-}}^{E_{d+}} (\varOmega_1,\dots,\varOmega_n)^T \in {\mathbb C}^{n}.
$$
A geodesic is doubly-periodic if and only if it intersects $Q\cap Q_d$ at a finite
number of complex points. In this case the linearized flow on ${\rm Jac}(\Gamma)$
must intersect $\Theta_{d\pm}$ at a finite set of points too.
In view of the Moser--Trubowitz isomorphism and Theorem \ref{N-section}, this holds if
$\Gamma$ is a hyperelliptic tangential cover of an elliptic curve ${\cal E}$.
Then, under the transformation (\ref{la-z}) with an appropriate $\beta$,
the real Weierstrass points on $\Gamma$
give real and positive parameters $a_i, c_j$ of the doubly-periodic geodesic.

Finally, there is an $(n-1)$-dimensional family of elliptic curves $\cal E$ in Jac($G$),
which is locally parameterized by points of their intersection with
the Abelian subvariety ${\cal A}_{n-1}$.
This gives rise to an $(n-1)$-dimensional family of the doubly-periodic geodesics. $\boxed{}$
\medskip

\paragraph{Remark.}
Since for any chosen $N$-fold tangential cover $G\mapsto {\cal E}$
the branch points of $G$ are functions of the two moduli $g_2,
g_3$, the parameters $a_i, c_j$ are uniquely determined by them
and by the rescaling factor $\alpha$ in (\ref{la-z}). This implies
that not any ellipsoid $Q$ may have doubly-periodic geodesics
associated {\it with the given degree of covering} as described by
Theorem \ref{KdV-Jacobi}. One can show that even in the simplest
case of a triaxial ellipsoid ($n=2$) and $N=3$ or 4, for any fixed
positive $a_1, a_2$ there exists only a finite number of possible
$c, a_3$ for which the geodesics are
doubly-periodic\footnote{Explicit algebraic conditions on
$a_1,a_2,a_3,c$ for the case of 3- and 4-fold tangential covers
were \\ presented in \cite{AF}.}.

Naturally, this does not exclude the existence of such geodesics
for other degrees of tangential coverings or those obtained from a
periodic flow on Jac($G$) via a birational transformation
different from (\ref{la-z}), or even just closed geodesics, which
are not doubly-periodic. However, the latter, if exist, cannot be
algebraic curves in view of the following property.

\begin{lemma} Any algebraic closed geodesic on ellipsoid $Q\subset {\mathbb R}^n$ is a
connected component of an elliptic or rational curve.
\end{lemma}

\noindent{\it Proof.} Let a closed geodesic be a connected component of an algebraic curve
$\cal C$. Since the geodesic flow on $Q$ is linearized on an unramified covering of Jac$(\Gamma)$,
$\cal C$ must be {\it an unramified} covering of an algebraic curve ${\cal C}_0\subset {\rm Jac}(\Gamma)$ and
and, moreover, ${\cal C}_0$ must be a one-dimensional Abelian subvariety.
Then, if $\Gamma$ is a regular curve and,
therefore,  Jac$(\Gamma)$ is compact, ${\cal C}_0$ can be only elliptic.
If $\Gamma$ has singularities (when, for example, $c_j=a_i$ and the geodesic lies completely
in hyperplane $X_i=0$) and its generalized Jacobian is not compact, then
${\cal C}_0$ can be also a rational curve.
In both cases $\cal C$, as an unramified covering of ${\cal C}_0$, can be only elliptic
or a reducible rational curve.  $\boxed{}$
\medskip


In the case $n=2$ the algebraic closed geodesics on a triaxial ellipsoid can
explicitly be expressed in terms of symmetric functions of the two ellipsoidal coordinates $\lambda_1, \lambda_2$ on $Q$.
 As a result, such geodesics can be rewritten
in terms of Cartesian coordinates in ${\mathbb R}^3$. We shall describe this procedure
in the next section.

\section{Genus 2 hyperelliptic tangential covers, algebraic polhodes,
and cutting algebraic surfaces in ${\mathbb R}^3$}

Suppose that the genus 2 hyperelliptic curve $G$
\begin{equation} \label{hyper}
w^{2}=-\prod_{k=1}^5 (z-b_k)
\end{equation}
is an $N$-fold tangential covering of the elliptic curve ${\cal
E}$ in (\ref{ell1}). Then, according to the Poincar\'e
reducibility theorem, $G$ is also an $N$-fold covering of another
elliptic curve
$$
{\cal E}_{2} =\left\{W^{2}=-\left(
4 Z^3-G_{2} Z -G_{3}\right) \equiv -4 (Z-E_1 )(Z-E_2)(Z-E_3)  \right\} ,
$$
the parameters $G_2, G_3$ being functions of the moduli $g_2, g_3$.

Let $U\in {\mathbb C}$ be uniformization parameter such that
$Z=\tilde\wp(U)$, $W=\tilde\wp'(U)$, and $\tilde\wp$ is the Weierstrass function
associated to the curve ${\cal E}_2$. As above, assume that the
point $\infty\in G$ is mapped to $U=0$. Then one can show that the map $\pi :\,
G\rightarrow {\cal E}_2$  is described by formulas
\begin{equation} \label{covs}
Z= {\cal Z}(z), \quad W= w^k \,{\cal W}(z),
\end{equation}
where $k$ is a positive odd integer number and 
${\cal Z}(z), {\cal W}(z)$ are rational functions of $z$ such that ${\cal Z}(\infty)=\infty$.
The second relation in (\ref{covs}) implies
that the Weierstrass points on $G$ are mapped to branch points on ${\cal E}_2$.

Consider the canonical embedding of $G$ to its Jacobian variety ${\mathbb C}^2=(u_1,u_2)/\Lambda$,
$$
P=(z,w)\mapsto {\cal A}(z,w)=\left( \int_{\infty}^{P}
\frac{dz}{w}, \int_{\infty}^{P} \frac{z\,dz}{w}\right)^T .
$$
The image of the embedding is the theta-divisor $\Theta$ that passes through the origin
in Jac$(G)$ and is tangent to vector ${\bf U}=(0,1)^T$.

The second covering $\pi :\, G\rightarrow {\cal E}_2$ is lifted to the Jacobian variety of
$G$. Namely, for any point ${\cal Q}\in {\cal E}_{2}$ and
$\pi ^{-1}({\cal Q})=\left\{P^{(1)},\dots,P^{(N)}\right\} \in G$, one has
\begin{equation} \label{*}
\int\limits_{\infty }^{P^{(i)}}\frac{dz}{w}=\kappa \int\limits_{\infty}^{\cal Q}\frac{d Z}{W},
\end{equation}
where $\kappa$ is a constant rational number depending on the degree $N$ only.
This implies that $z$-coordinates of the $N$ points of intersection of a complex $u_2$-line ($u_1=$const)
with $G=\Theta\subset\mbox{Jac}(G)$ are the roots of the first equation in (\ref{covs})
with $Z=Z({\cal Q})$ (see also \cite{Smirnov2}).


Now let $P_{1}=(z_{1},w_{1}),P_{2}=(z_{1},w_{2})\in G$ and consider the full Abel--Jacobi map
\begin{equation} \label{AJ}
{\cal A}(P_1)+{\cal A}(P_2)=(u_1, u_2)^T .
\end{equation}
Assume that $u_1, u_2$ evolve according to ${\bf U}$-flow, that is $u_1=$const. Hence $z_1, z_2$
satisfy the equations
\begin{equation} \label{eqn-z}
\dot z_1=\frac{w_1}{z_1-z_2}, \quad \dot z_2=\frac{w_1}{z_2-z_1}.
\end{equation}
This imposes a relation between coordinates of $P_1$ and $P_2$ on
$G$. In the generic case, the relation is transcendental one and
the coordinates are quasiperiodic functions of time. However, if
$G$ is a tangential covering of an elliptic curve, then the
relation becomes algebraic and can be found explicitly in each
case of covering. Namely, let us set
\begin{equation} \label{sum}
U_1=\int\limits_{\infty}^{\pi(P_1)} \frac{dZ}{W}, \quad 
U_2=\int\limits_{\infty}^{\pi(P_2)}  \frac{dZ}{W}, 
\quad\mbox{and}\quad U_*=U_1+U_2 .
\end{equation}
In view of (\ref{*}) and the condition $u_1=$const, the first equation
in (\ref{AJ}) implies $U_*=$const.

Next, due to the addition theorem for elliptic functions,
\begin{equation} \label{add_diff} \left\vert
\begin{array}{ccc}
1 & \tilde\wp (U_{1}) & \tilde\wp ^{\prime }(U_{1}) \\
1 & \tilde\wp (U_{2}) & \tilde\wp ^{\prime }(U_{2}) \\
1 & \tilde\wp (-U_*) & \tilde\wp ^{\prime }(-U_*)
\end{array}
\right\vert =0,
\end{equation}
or, in the integral form
$$
\tilde\wp(U_*) +\tilde\wp(U_1)+ \tilde\wp(U_2)=
- \frac 14 \left[\frac{ \tilde\wp'(U_1)-\tilde\wp'(U_2)}{ \tilde\wp(U_1)-\tilde\wp(U_2) } \right]^2,
$$
the coordinates $Z_1=\tilde\wp(U_1), Z_2=\tilde\wp(U_2)$ are subject to the constraint
\begin{align*}
2G_{3}+G_{2}\left( Z_{1}+Z_{2}\right) & +4 \wp(U_*) \left(
Z_{2}-Z_{1}\right)^{2}  -4 Z_{2}Z_{1}\left( Z_{1}+Z_{2}\right) \\
&\quad = -2 \sqrt{ 4 Z_{1}^{3}-G_{2} Z_{1}-G_{3}} \sqrt{4Z_{2}^{3}-G_{2} Z_{2}-G_{3}}.
\end{align*}
Then, taking square of both sides, simplifying, factoring out $(Z_1-Z_2)^2$, and
replacing $Z_1, Z_2$ by the expressions ${\cal Z}(z_1), {\cal Z}(z_2)$ from (\ref{covs}), we arrive at
{\it generating equation\/}
\begin{align}
16 & \left({\cal Z}(z_2)-{\cal Z}(z_1) \right)^{2} \widetilde\wp_*^{2} \nonumber  \\
& +\left[ 16G_{3}+8G_{2}\left({\cal Z}(z_1)  + {\cal Z}(z_2) \right)
 -32{\cal Z}(z_1) {\cal Z}(z_2) ({\cal Z}(z_1) + {\cal Z}(z_2) ) \right ] \widetilde\wp_* \nonumber \\
& + 16G_{3} \left({\cal Z}(z_1) + {\cal Z}(z_2)\right)
+8G_{2} {\cal Z}(z_1) {\cal Z}(z_2) +16 {\cal Z}(z_1)^{2} {\cal Z}(z_2)^{2} +G_{2}^{2}=0.
\label{main}
\end{align}
Written in terms of $Z_1,Z_2$, it 
defines an elliptic curve isomorphic to  ${\cal E}_2$ for any $\widetilde\wp_*$.

In terms of $z_1,z_2$, the generating equation gives a family of
algebraic curves ${\cal H}_{\tilde\wp_*} \subset (z_1,z_2)$, which we call {\it polhodes}.
They are symmetric with respect to the diagonal $z_1=z_2$, as expected, and, for a generic
$\tilde\wp_*$, has degree $4N$\footnote{
However, as seen from the structure of (\ref{main}), a fixed generic $z_1$ (and $u_1$)
results in $2N$ (complex) solutions $z_2$.}.
A polhode describes an algebraic relation between $z$-coordinates of the divisor
$P_1=(z_{1},w_{1}), P_2=(z_{1},w_{2})$ on $G$, which holds under the $u_{2}$-flow on Jac$(G)$.
The parameter $\widetilde\wp_*=\tilde\wp(U_*)$ plays the role of a constant phase of the flow.
\medskip

The polhodes thus can be regarded as ramified coverings of  ${\cal E}_2$ and, therefore,
in general, have genus $>1$.


\paragraph{Real finite asymmetric part of polhodes.}
Suppose that all the roots of the degree 5 polynomial in (\ref{hyper}) are real and set
\begin{equation} \label{BS}
b_1 < b_2 <\cdots < b_5.
\end{equation}
Assume that the variables $z_1, z_2$ range in finite segments $[b_i, b_j]$, where
both $w_1, w_2$ are real and finite.
Taking in mind applications to problems of dynamics,
we also assume that these segments are different and $z_1<z_2$.
Then the motion of the point $(z_1, z_2)$ is bounded in the unique square domain
$$
S = \{ b_2\le z_1 \le b_3, \;  b_4\le z_2 \le b_5\}.
$$
Let also $\tilde\wp_*$ in (\ref{main}) be real.
The part of polhode ${\cal H}_{\tilde\wp_*}\subset (z_1, z_2)$ that lies in $S$
will be called the {\it real asymmetric part} of ${\cal H}_{\tilde\wp_*}$.
At the vertices of the domain both $w_1, w_2$ equal zero. Then,
in view of equations (\ref{eqn-z}), this part of the polhode
is tangent to the sides of $S$ or passes through some of its vertices.

\begin{lemma} \label{reality}
If $U_*$ is such that in the domain $S$ one of the following relations holds
\begin{equation} \label{roots}
\tilde\wp (U_*)= {\cal Z}(z_1), \quad \mbox{or} \quad \tilde\wp (U_*)={\cal Z}(z_2),
\end{equation}
then the real asymmetric part of ${\cal H}_{\tilde\wp_*} $ is empty.
\end{lemma}

\noindent{\it Proof.} Indeed, in view of (\ref{sum}), condition $\tilde\wp_*= Z(z_1)$ implies
$U_2=0$. Hence, for the above value of $z_1\in [b_2, b_3]$, the coordinate
$z_2$ must be infinite.
If the component of ${\cal H}_{\tilde\wp_*}$ in $S$ is not empty, then the polhode must intersect
the boundary of $S$, which is not possible. Hence this component is empty.

If the second condition in (\ref{roots}) is satisfied, the proof goes along similar lines.
$\boxed{}$
\medskip

In view of the above lemma, we also assume that the constant
parameter $\tilde\wp_*$ lies in a segment on $\mathbb R$ where neither
of the conditions (\ref{roots}) is satisfied, which is one of the
gaps $[E_\alpha,E_\beta]$, $[-\infty,E_1]$, $[E_3,\infty]$.

\paragraph{Special polhodes.} If the parameter $\tilde\wp_*$ in (\ref{main}) coincides with
a branch point of ${\cal E}_2$, then the equation of the polhode simplifies.

In the first obvious case $\tilde\wp_*=\infty$ the generating equation (\ref{main}) reduces to
$Z(z_1)-Z(z_2)=0$. Since $Z(z)$ is a rational function, from here
one can always factor out $z_1-z_2$.
Thus, the connected component of the polhode in the domain $S$ is
\begin{equation} \label{infty}
{\cal H}_\infty =\left\{\frac{Z(z_1)-Z(z_2)}{z_1-z_2}=0   \right\} .
\end{equation}
\medskip

Next, for $\tilde\wp(U_*)=E_\alpha$, $\tilde\wp'(U_*)=0$,
from the addition formula (\ref{add_diff}) we obtain
the following simple equation
\begin{equation} \label{simple}
\tilde\wp'(U_1)(E_\alpha-\tilde\wp(U_2)) =\tilde\wp'(U_2)(E_\alpha-\tilde\wp(U_1))) .
\end{equation}
Taking squares of both sides and simplifying, we get
\begin{gather*}
4 (Z_1-Z_2)(E_\alpha-Z_2) (E_\alpha-Z_1)
\cdot(E_\alpha (Z_1+Z_2) -Z_1 Z_2 +E_\alpha^2+E_\beta E_\gamma)=0 ,\\
 (\alpha,\beta,\gamma)=(1,2,3) .
\end{gather*}
Then we factor out the term $(Z_1-Z_2)$ that leads to polhode
${\cal H}_\infty$, as well as the product $(E_\alpha-Z_2)(E_\alpha-Z_1)$
that leads to two lines in $(z_1, z_2)$-plane and therefore cannot
describe the polhode. As a result, we obtain the {\it special generating equation}
\begin{equation} \label{sp_gen}
({\cal Z}(z_1)-E_\alpha)({\cal Z}(z_2)-E_\alpha ) -2 E_\alpha^2 - E_\beta E_\gamma=0 ,
\end{equation}
which defines the special polhode ${\cal H}_{E_\alpha}$.
For a fixed generic $z_1$, this equation has $N$ complex solutions for $z_2$.

\begin{lemma}
The polhodes ${\cal H}_\infty$, ${\cal H}_{E_\alpha}$ pass through
two vertices of the domain $S$.
\end{lemma}

\noindent{\it Proof.} Since 6 branch points of $G$ are mapped to 4
branch points of ${\cal E}_2$, some different finite branch points
of $G$ are mapped to the same finite branch point on the elliptic
curve. Thus, at two vertices of $S$, $ {\mathcal Z}(z_1)={\mathcal
Z}(z_2)$ for $z_1\ne z_2$ and the polhode ${\cal H}_\infty$ passes
through these vertices.

Next, at the vertices of $S$ one has $w_1=w_2=0$, and,
in view of the second relation in (\ref{covs}),
$\tilde\wp'(U_1)=\tilde\wp'(U_2)=0$. Hence equation (\ref{simple})
is satisfied in all the vertices. On the other hand, at two of the four vertices
the condition $(E_\alpha-u_2)(E_\alpha-u_1)=0$ is also satisfied. Since the
product  $(E_\alpha-u_2)(E_\alpha-u_1)$ was factored out in (\ref{sp_gen}),
the polhode  ${\cal H}_{E_\alpha}$ does not pass through the latter two vertices,
hence it passes through the other two.


\paragraph{Polhodes and Closed Geodesics on an Ellipsoid.}
Let $(\beta,0)$ be a finite Weierstrass point on $G$.
Then under the birational transformation $(z,w)\mapsto (\lambda,\mu)$ given by (\ref{la-z})
with $\alpha=1$ and $\beta=b_1$ (the minimal root of (\ref{BS}))
the curve $G$ passes to a genus 2 curve
$$
\Gamma=\{\mu^2=-\lambda( \lambda-a_1) (\lambda-a_2) (\lambda-a_3) (\lambda-c)\},
$$
such that $a_i$ and $c$ are {\it positive}.
Thus $\Gamma$ can be regarded as the spectral curve of the geodesic flow on the ellipsoid
$$
Q= \left\{ \frac{X_1^2}{a_1}+ \frac{X_2^2}{a_2}+ \frac{X_3^2} {a_3}=1 \right\} ,
\qquad a_1<a_2<a_3
$$
and the corresponding variables
\begin{equation} \label{LAZ}
\lambda _{1}=\frac{1}{z_{1}-b_1}, \quad \lambda _{2}=\frac{1}{z_{2}-b_1},
\end{equation}
are the ellipsoidal coordinates of the moving point on $Q$.

In view of Theorem \ref{KdV-Jacobi}, under the transformation
(\ref{la-z}) the real asymmetric part of polhode ${\cal
H}_{\tilde\wp_*}$ describes a closed geodesic on the ellipsoid $Q$ in
terms of the ellipsoidal coordinates,
whereas the whole family of the polhodes gives a one-parametric family
of such geodesics that are tangent to one and the same caustic on $Q$.

Substituting expressions $z_1(\lambda_1), z_2(\lambda_2)$ into the
generating equation (\ref{main}), one obtains equation of the
geodesic in terms of symmetric functions
$\Sigma_1=\lambda_1+\lambda_2$, $\Sigma_2=\lambda_1\lambda_2$ of degree $2N$. In
view of relations (\ref{spheroconic2}) for $n=2$, the latter can
be expressed via the Cartesian coordinates as follows
\begin{equation}
\label{spheroconic}
\begin{aligned}
\Sigma_1 &= a_1+a_3 + \frac 1{a_1}(a_2-a_1) X_1^2 + \frac 1{a_3}(a_2-a_3) X_3^2, \\
\Sigma_2 &= a_1 a_3 + \frac{a_3}{a_1}(a_2-a_1) X_1^2 + \frac{a_1}{a_3}(a_2-a_3) X_3^2 .
\end{aligned}
\end{equation}
As a result, one arrives at equation of an algebraic cylinder
surface ${\cal V}_{\tilde\wp_*}$ of degree 4N in ${\mathbb R}^3$,
which cuts out a closed geodesic on $Q$. More precisely, one get a
family of such surfaces parameterized by $\tilde\wp_*$.

\paragraph{Remark.}
Since the equation depends on squares of $X_i$ only,
such surfaces are symmetric with respect to reflections $X_i \to -X_i$. Thus, the
complete intersection ${\cal V}_{\tilde\wp_*}\cap Q$ consists of {\it a union\/} of closed geodesics
that are transformed to each other by these reflections. An example of such intersection
is given in Figure \ref{sp1.fig}.

As we shall see below, in some cases the equation admit a factorization
and the cylinder ${\cal V}_{\tilde\wp_*}$ splits in two connected non-symmetric components.
\medskip

It should be emphasized that the method of polhodes is based on the existence of 
the second covering $G\mapsto {\mathcal E}_2$ and the addition law on ${\mathcal E}_2$, so
it does not admit a straightforward generalization to a similar description of 
algebraic closed geodesics on $n$-dimensional ellipsoids ($n>2$). 
Indeed, as mentioned in Section 3, 
in this case ${\mathcal E}_2$ is replaced by an Abelian subvariety 
${\mathcal A}_{n-1}$, for which an algebraic description is not known.
\medskip

In the sequel we consider in detail polhodes ${\cal H}_{\tilde\wp_*}$
and surfaces ${\cal V}_{\tilde\wp_*}$ for the 3:1 and 4:1 hyperelliptic
tangential covers.

\section{The 3:1 tangential cover (the Hermite case)}
In this case first indicated by Hermite (\cite{Herm}, see also \cite{Enol, Smirnov2}),
the elliptic curve ${\cal E}_1$ in
(\ref{ell1}) is covered by the genus 2 curve
\begin{equation} \label{hyper3}
G=\left\{w^{2}=-\frac{1}{4}\left ( 4z^{3}-9g_{2}z-27g_{3}\right) (z^{2}-3g_{2})\right\} .
\end{equation}
The latter also covers the second elliptic curve
\begin{gather}
{\cal E}_{2} =\left\{ W^{2} =-\left(
4 W^{3}-G_{2} W -G_{3}\right)\equiv -4 (Z-E_1 )(Z-E_2)(Z-E_3)\right\} , \label{3-cover} \\
G_{2} = \frac{27}{4}\left( g_{2}^{3}+9g_{3}^{2}\right) ,\quad
G_{3}=\frac{243}{8}g_{3}\left( 3g_{3}^{2}-g_{2}^{3}\right) \label{G's}
\end{gather}
and the covering formulas (\ref{covs}) take the form
\begin{equation} \label{cov}
Z =\frac{1}{4}\left( 4z^{3}-9g_{2}z-9g_{3}\right) , \quad
W =-\frac{w}{2}\left( 4z^{2}-3g_{2}\right) .
\end{equation}
The roots of the polynomial in (\ref{hyper3}) are real iff $g_2, g_3$ are real and
$g_2>0$, $27 g_3^3> g_2^3$. Then, assuming that $e_1<e_2<e_3$, $E_1<E_2<E_3$,
the following ordering holds
\begin{gather}
b_1= - \sqrt{3 g_2}, \quad
\{b_2, b_3, b_4 \}= \{3e_1, 3e_2, 3e_3\}, \quad b_5=\sqrt{3 g_2}, \nonumber \\
E_1= -\frac 34 \left(3 g_3 + \sqrt{3 g_2^3} \right), \quad
E_2=-\frac 34 \left(3 g_3- \sqrt{3 g_2^3}\right), \quad E_3=\frac 92 g_3, \label{order-3}
\end{gather}
and
\begin{equation} \label{corr-3}
Z(b_1)=E_1, \quad Z(b_5)=E_2, \quad Z(b_2)=Z(b_3)=Z(b_4)= E_3.
\end{equation}

Now substituting (\ref{cov}) into the generating equation
(\ref{main}) and taking into account (\ref{G's}), we get following
family of polhodes
\begin{equation} \label{final-3}
 M_2 (z_1,z_2) \,\tilde\wp_*^2+ M_1(z_1,z_2)\, \tilde\wp_* + M_0(z_1,z_2)  =0 ,
\end{equation}
where
\begin{align*}
M_{2} & =\frac{1}{16}(z_{2}-z_{1})^{2}\, (9g_{2}-4z_{1}z_{2}-4z_{1}^{2}-4z_{2}^{2})^{2}, \\
M_{1} & = -\frac{729}{16}g_{2}^{3}g_{3} -\frac{243}{32}g_{2}^{4}\left( z_{1}+z_{2}\right)
-2z_{2}^{3}z_{1}^{3}\left( z_{1}^{2}-z_{1}z_{2}+z_{2}^{2}\right) (z_{1}+z_{2}) \\
& +\frac{9}{2} g_2\,
\left(z_{1}^{4}+z_{2}^{4}-z_{1}z_{2}^{3}-z_{1}^{3}z_{2}+3z_{1}^{2}z_{2}^{2}\right)
+\frac{729}{32} g_{3}g_{2}^{2}\, \left( 4z_{1}z_{2}+z_{1}^{2}+z_{2}^{2}\right)
(z_{1}+z_{2}) z_{1}z_{2} \\
& -\frac{81}{4}g_{3} g_{2}\left(z_{1}^{4}+z_{2}^{4}+2z_{1}z_{2}^{3}+2z_{1}^{3}z_{2}\right)
+\frac{27}{32}\,g_{2}^{3} \left( 23z_{1}z_{2}+4z_{1}^{2}+4z_{2}^{2}\right)
(z_{1}+z_{2}) \\
& -\frac{81}{8} g_{2}^{2}\, \left(2z_{1}^{2}-z_{1}z_{2}+2z_{2}^{2}\right)(z_{1}+z_{2})
z_{1}z_{2} +\frac{9}{2}g_{3}\, \left(z_{1}^{6}+z_{2}^{6}+4z_{1}^{3}z_{2}^{3}\right) , \\
M_0 & = \frac{1}{256}\left(729g_{2}^{6}+4374g_{2}^{5}z_{1}z_{2}
+52\,488g_{2}^{3}g_{3}^{2}+256z_{1}^{6}z_{2}^{6}\right) \\
& +\frac{10\,935}{128}g_{3}g_{2}^{4}\left( z_{1}+z_{2}\right)
-\frac{9}{2} g_{3}\left( z_{1}+z_{2}\right) \left( z_{1}^{2}-z_{1}z_{2}+z_{2}^{2}\right)
z_{1}^{3}z_{2}^{3} \\
& -\frac{9}{2}g_{2}\left(z_{1}^{2}+z_{2}^{2}\right) z_{1}^{4}z_{2}^{4}
+ \frac{6561}{256}g_{3}^{2}g_{2}^{2} \left( 10z_{1}z_{2}+z_{1}^{2}+z_{2}^{2}\right) \\
& +\frac{81}{16}g_{3}^{2}\left(z_{1}^{6}+z_{2}^{6}+10z_{1}^{3}z_{2}^{3}\right)
-\frac{243}{256} g_{2}^{4}\left( 8z_{1}^{2}-27z_{1}z_{2}+8z_{2}^{2}\right) z_{1}z_{2} \\
& +\frac{81}{16}g_{2}^{2}\left( z_{1}^{4}+z_{2}^{4}+4z_{1}^{2}z_{2}^{2}\right)
z_{1}^{2}z_{2}^{2} -\frac{27}{32}g_{2}^{3}\left(
27z_{1}^{2}-4z_{1}z_{2}+27z_{2}^{2}\right) z_{1}^{2}z_{2}^{2} \\
& -\frac{729}{32}g_{3}^{2} g_{2}
\left(z_{1}^{4}+z_{2}^{4}+5z_{1}z_{2}^{3}+5z_{1}^{3}z_{2}\right)
 -\frac{243}{128}g_{3}g_{2}^{3}\left(
20z_{1}^{2}-47z_{1}z_{2}+20z_{2}^{2}\right) \left( z_{1}+z_{2}\right) \\
&+ \frac{81}{8}g_{3}g_{2}
\left(z_{1}^{4}+z_{2}^{4}-z_{1}z_{2}^{3}-z_{1}^{3}z_{2}+3z_{1}^{2}z_{2}^{2}\right)
(z_{1}+z_{2})z_{1}z_{2} \\
& -\frac{729}{32}g_{3}g_{2}^{2}\left( 2z_{1}^{2}-z_{1}z_{2}+2z_{2}^{2}\right)
(z_{1}+z_{2}) z_{1} z_{2}.
\end{align*}

\paragraph{Reality conditions.} Assume that the parameter $\tilde\wp_*$ is real and
$(z_1,z_2)$ range in the square domain $S$, namely
$$
S = \{ 3e_1 \le z_1 \le 3e_2, \;  3e_3 \le z_2 \le \sqrt{3g_2} \}.
$$
 Then, from relations (\ref{order-3}),
(\ref{corr-3}), we conclude that the conditions (\ref{roots}) in Lemma \ref{reality}
cannot be satisfied if and only if $\tilde\wp_*$ varies in the gap
$(-\infty; E_1]$. 

\paragraph{Limit Polhodes.}
When $\tilde\wp_*=\infty$, equation (\ref{final-3}) 
reduces to
\begin{equation} \label{quadr}
 9g_{2}-4z_{1}z_{2}-4z_{1}^{2}-4z_{2}^{2}=0 ,
\end{equation}
which can also be obtained directly from (\ref{infty}). Thus ${\cal H}_\infty$
defines a conic $\cal C$ in $(z_1,z_2)$-plane, which passes through 2 vertices of the
domain $S$.
\medskip

Next, setting in (\ref{sp_gen}) $\tilde\wp_*=E_1=-\frac 34 (3 g_3- \sqrt{3 g_2^3})$ and
taking into account expressions (\ref{G's}) results in cubic equation
\begin{align}
& 16z_{1}^{3}z_{2}^{3}-12\sqrt{3 g_2^3} \left( z_{1}^{3}+z_{2}^{3}\right)
-36g_{2}z_{2}z_{1}\left( z_{1}^{2}+z_{2}^{2}\right) \nonumber \\
 &\quad  +81g_{2}^{2}z_{1}z_{2}+27g_{2}\sqrt{3 g_2^3} \left( z_{1}+z_{2}\right)
+162\sqrt{3 g_2^3} g_{3}-27g_{2}^{3}=0 . \label{E_1}
\end{align}
The corresponding polhode ${\cal H}_{E_1}$ also passes through two vertices of $S$.

\paragraph{Examples of Polhodes in $S$.}
To illustrate the above polhodes, in (\ref{ell1}) we choose
$$
g_{2}= 3, \quad g_{3}=0.2 .
$$
Then the roots of the polynomials
$\frac{1}{4}\left( 4z^{3}-9g_{2}z-27g_{3}\right)(z^{2}-3g_{2})$ and \\
$4 Z^{3}-G_{2} Z -G_{3}$ are, respectively,
\begin{equation}\label{ex1}
(3.0, \; 2.693, \; -0.201, \; -2.492, \; -3.0),
\quad \mbox{and} \quad (6.3, \; 0.9, \; -7.2) .
\end{equation}
The domain $S$, where the corresponding variables $w_1, w_2$ are real, is
\begin{equation}\label{real}
S= \{ -2.492\le z_1 \le-0.201, \quad 2.693\le z_2 \le 3.0 \} .
\end{equation}

If the parameter $\tilde\wp_*$ varies in  the interval $(-\infty;-7.2)$, then
the real roots of the equation
$$
\tilde\wp_* =\frac{1}{4}\left( 4z^{3}-9g_{2}z-9g_{3}\right)
$$
do not fit into the intervals in (\ref{real}). This means that the
conditions (\ref{roots}) do not hold in domain $S$ and the real
asymmetric part of the polhode may be non-empty.

Note that if $\tilde\wp_*$ belongs to other intervals on the real line,
some of these conditions are necessarily satisfied, so we exclude
this case from consideration.

\begin{figure}[hb]
\begin{center}
\subfigure[$\tilde\wp_* =-8.0$]{
\includegraphics[height=0.25\textwidth]{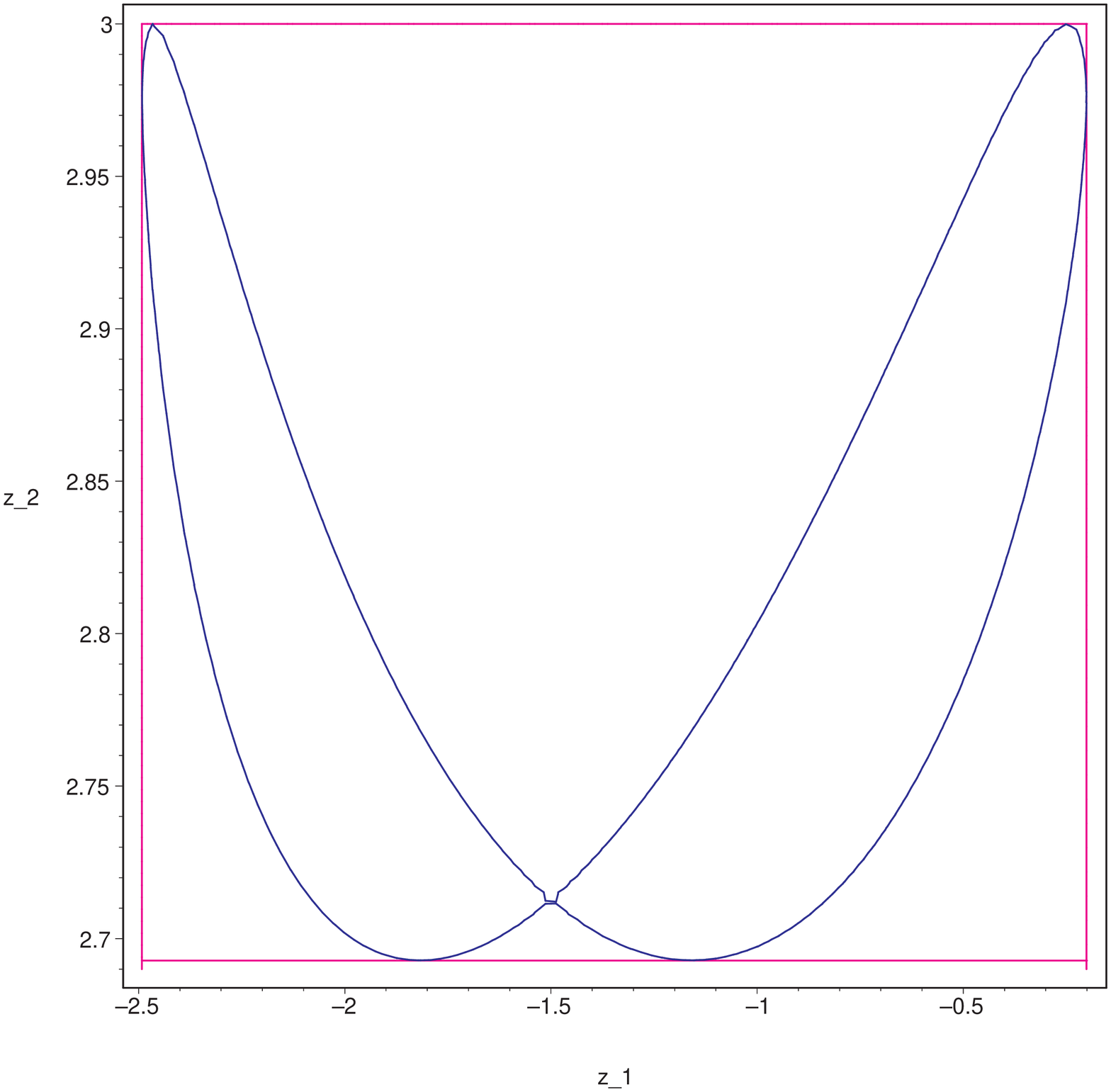}} \quad
\subfigure[$\tilde\wp_* =-11.0$]{
\includegraphics[height=0.25\textwidth]{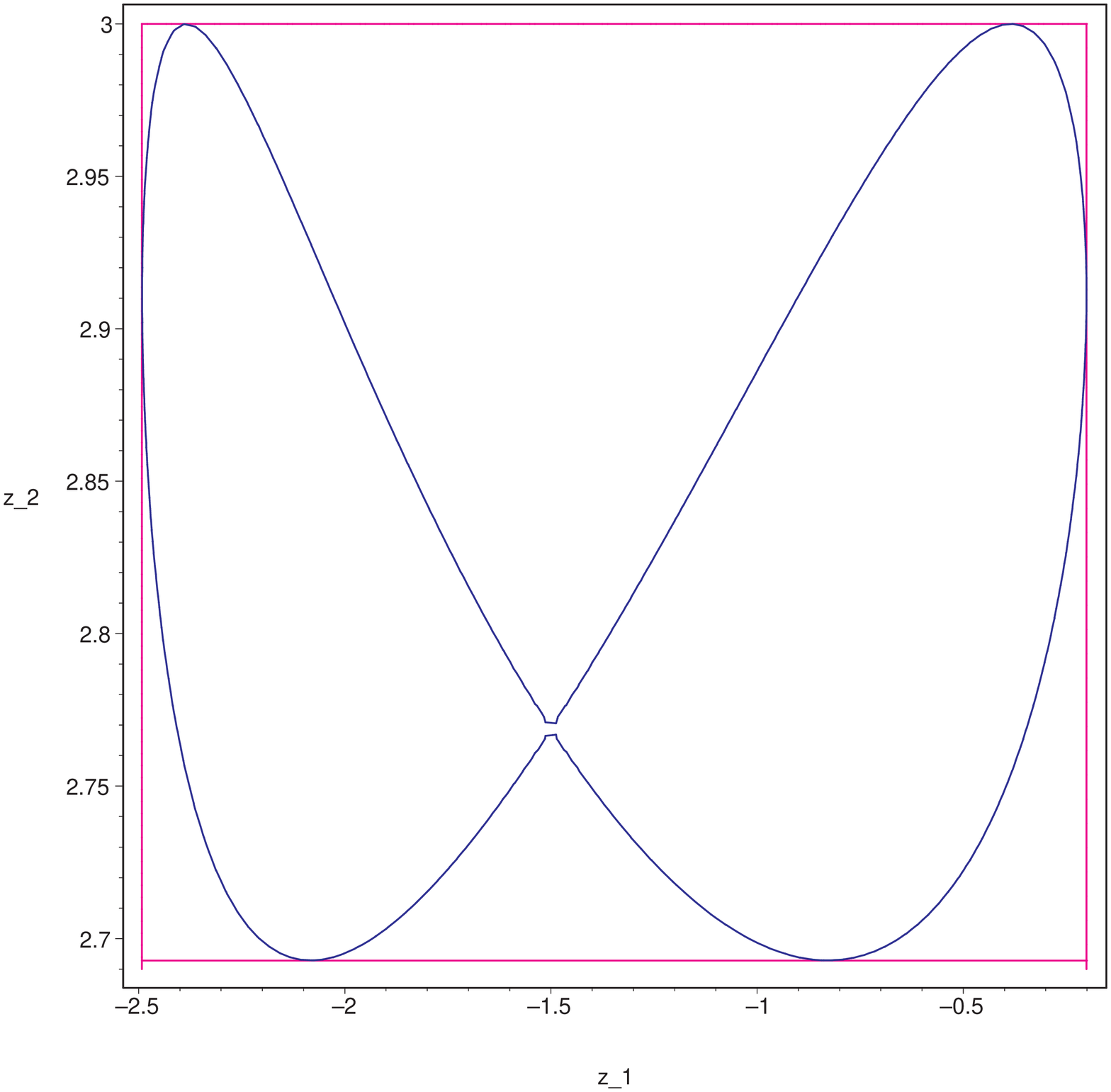}}
\end{center}
\caption{Generic polhodes in $S\subset {\mathbb R}^2=(z_1, z_2)$.}\label{polodias3-1.fig}
\end{figure}

\begin{figure}[hb]
\begin{center}
\subfigure[$\tilde\wp_* =-\infty$]{
\includegraphics[height=0.25\textwidth]{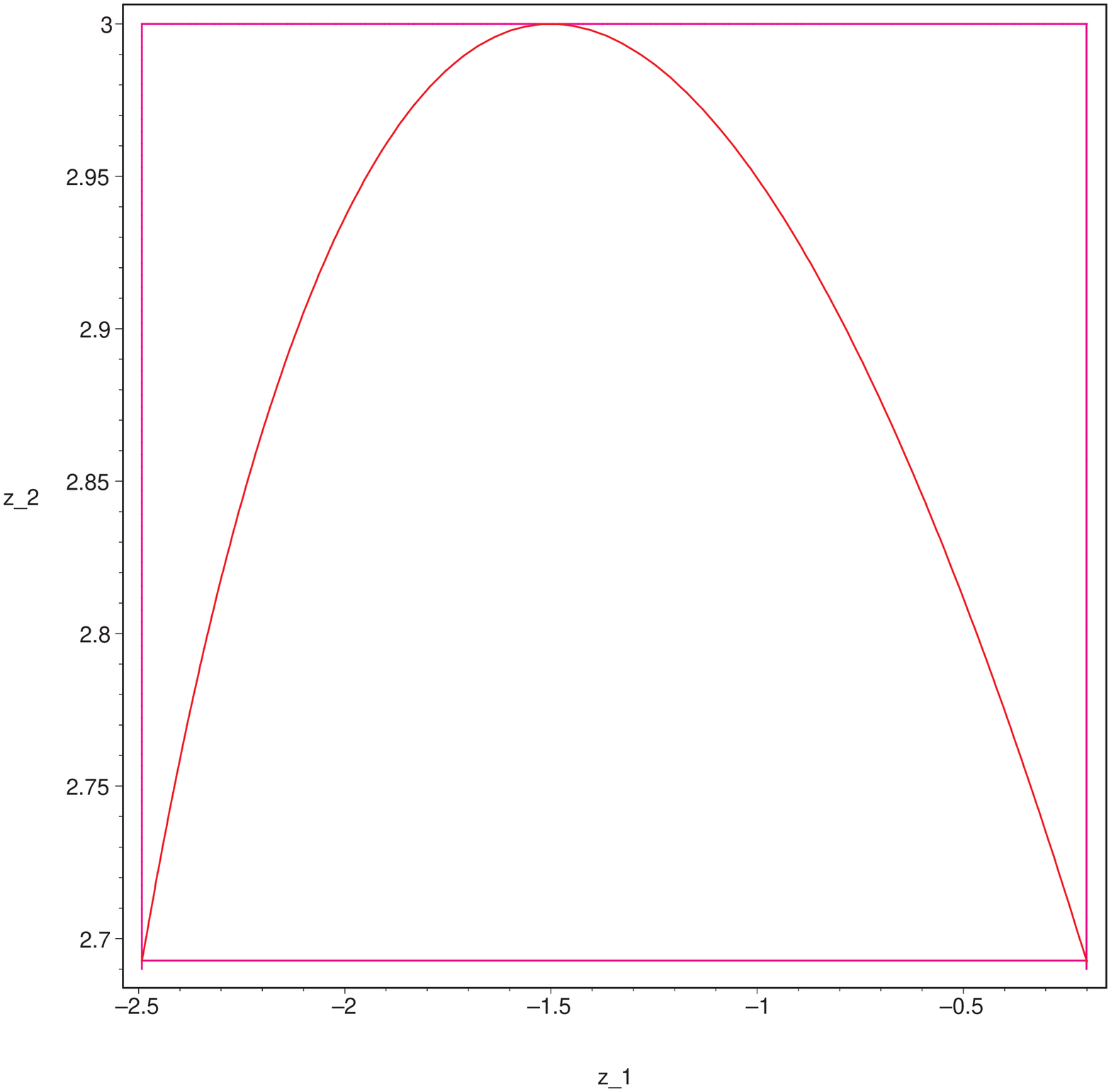}} \quad
\subfigure[$\tilde\wp_* =E_1=-7.2$]{
\includegraphics[height=0.25\textwidth]{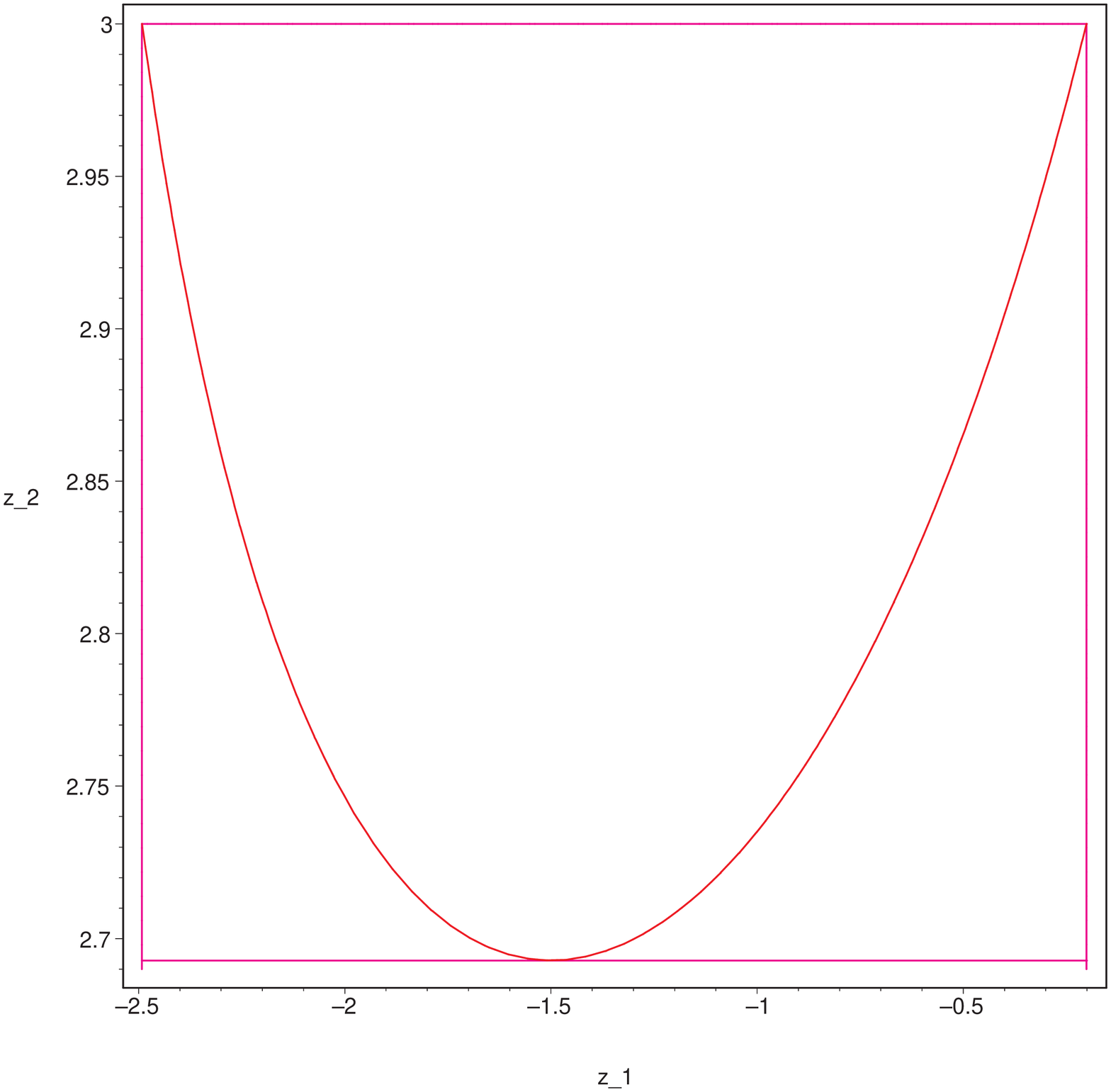}}
\end{center}
\caption{Special polhodes in $S$.}\label{limit.polodias.fig}
\end{figure}

The graphs of equation (\ref{final-3}) in the domain (\ref{real}) for two generic
values of $\tilde\wp_*$ are given in Figure \ref{polodias3-1.fig}, whereas
the graphs of the special polhode ${\cal H}_\infty$ given by equation (\ref{quadr}) and
${\cal H}_{E_1}$ given by (\ref{E_1}) are presented in Figure \ref{limit.polodias.fig}.

As seen from Figure \ref{polodias3-1.fig}, generic polhodes in $S$
intersect generic lines $z_2=$const and $z_1=$const at 4 and 2 points respectively.
\newpage

\paragraph{Closed geodesics related to 3:1 covering.}
Under the  the projective transformation (\ref{la-z}) with
$\beta=b_1= -\sqrt{3g_2}$, the branch points $\{-\sqrt{3g_2}, 3e_1, 3e_2, 3e_3, \sqrt{3g_2}, \infty\}$ of the curve
$G$ transform to infinity, 4 positive numbers $\{a_1, a_2,a_3,c\}$ and zero respectively.
Given $g_2, e_1$, the parameters $e_2,e_3$ of the
elliptic curve are defined uniquely:
$$
e_2 = -\frac {e_1} 2 -\frac{\sqrt{3}}{6} R, \quad
e_3 = -\frac {e_1} 2 +\frac{\sqrt{3}}{6} R, \qquad R=\sqrt{3g_2-9 e_1^2}.
$$
Then, assuming that $a_1< a_2<c<a_3$, we get
\begin{align}
 a_3 & = \frac 1{3e_1+B},\quad a_1=\frac 1{2B}, \nonumber \\
a_2 &= \frac 1{3e_3+B}\equiv \frac {2}{(6e_3-B)^2} \left(2B-3e_3 -\sqrt{3}R\right),
\label{choice}\\
c & =\frac 1{3e_2+B} \equiv \frac {2}{(6e_3-B)^2} \left(2B-3e_3 +\sqrt{3}R \right),
\nonumber
\end{align}
where $B=\sqrt{3g_2}$.
As a result, the four parameters $a_1< a_2<c<a_3$ are uniquely defined by $g_2$ (or $B$)
and $e_1$.
\medskip

Now we apply the transformation (\ref{LAZ}) with
$b_1= -\sqrt{3g_2}$ to the polhode (\ref{final-3}).
This yields an equation of a closed geodesic on $Q$ written in terms of the symmetric functions
$\Sigma_1=\lambda_1+\lambda_2$, $\Sigma_2=\lambda_1\lambda_2$ of the
ellipsoidal coordinates. (In fact, one obtains a family of such geodesics parameterized by
$\tilde\wp_*$.) Then, making the substitution (\ref{spheroconic})
one obtains the equation of the cylinder cutting surface
${\cal V}_{\tilde\wp_*}$ in terms of squares of the Cartesian coordinates
$X_1,X_3$. For a generic parameter $\tilde\wp_*$ this equation has
degree 12, it is quite tedious and we do not give it here.
However, the structure of a generic polhode in $S\subset
(z_1,z_2)$ and the correspondence between the sets $\{3e_1, 3e_2,3e_3, \sqrt{3g_2}\}$
and $\{a_1,a_2,c,a_3\}$ is already sufficient to give a complete
qualitative description of the geodesic on $Q$.

Namely, let ${\cal R}_c$ be a ring on $Q$
bounded by the two connected components of the caustic $Q\cap Q_c$ and $\rho=k:l\in{\mathbb Q}$
be the quotient
of the numbers of complete rotations performed by a closed geodesics in lateral and meridional
directions on the ring respectively (the rotation number).

\begin{theorem} \label{2-types} \begin{description}
\item{1).} Under the assumption $a_1< a_2<c<a_3$, the geodesic corresponding to a generic
polhode (\ref{final-3}) or to the special polhodes 
is located in the ring ${\cal R}_c$ between planes
$X_3=\pm h$, $h<\sqrt{a_3}$ and has rotation number $\rho=2:1$.
It touches the caustic $Q\cap Q_c$ at 2 points and has one self-intersection.

\item{2).} Under the assumption $a_1< c< a_2 <a_3$, the geodesic
is located in the ring ${\cal R}_c$  between planes $X_1= \pm h$, $h<\sqrt{a_1}$ and
has rotation number
$1:2$. It touches the caustic $Q\cap Q_c$ at 4 points and has no self-intersections.
\end{description}
In both cases the geodesic is either a 2-fold covering of the real
asymmetric part of a generic polhode ${\cal H}_{\tilde\wp_*}$ or a
4-fold covering of that of the special polhodes.
\end{theorem}

Note that the self-intersection point of the polhode {\it does not} correspond to the
self-intersection point of the corresponding closed geodesic.
\medskip


\noindent{\it Sketch of Proof of Theorem} \ref{2-types}. Under the projective
transformation (\ref{LAZ}), a polhode ${\cal H}_{\tilde\wp_*}\subset (z_1, z_2)$
is mapped to a polhode
$\widetilde{\cal H}_{\tilde\wp_*}$ in ${\mathbb R}^2=(\lambda_1,\lambda_2)$, which is
tangent to lines $\lambda_j=a_i$, $i=1,2,3$ and $\lambda_j=c$.
In view of relations (\ref{spheroconic2}), 
the point of tangency of $\widetilde{\cal H}_{\tilde\wp_*}$
to the line $\lambda_j=a_i$ corresponds to the moment when
the geodesic $X(s)$ on $Q$ crosses the plane $X_i=0$, and the tangency to
the line $\lambda_j=c$ corresponds to the tangency of $X(s)$ to the caustic $Q\cap Q_c$.
Estimating ordering and number of the tangencies of $\widetilde{\cal H}_{\tilde\wp_*}$ in the
cases $a_2<c$ and $c<a_2$, one arrives at the statements of the theorem. $\boxed{}$

\begin{figure}[float]
\begin{center}
{\includegraphics[height=0.75\textwidth]{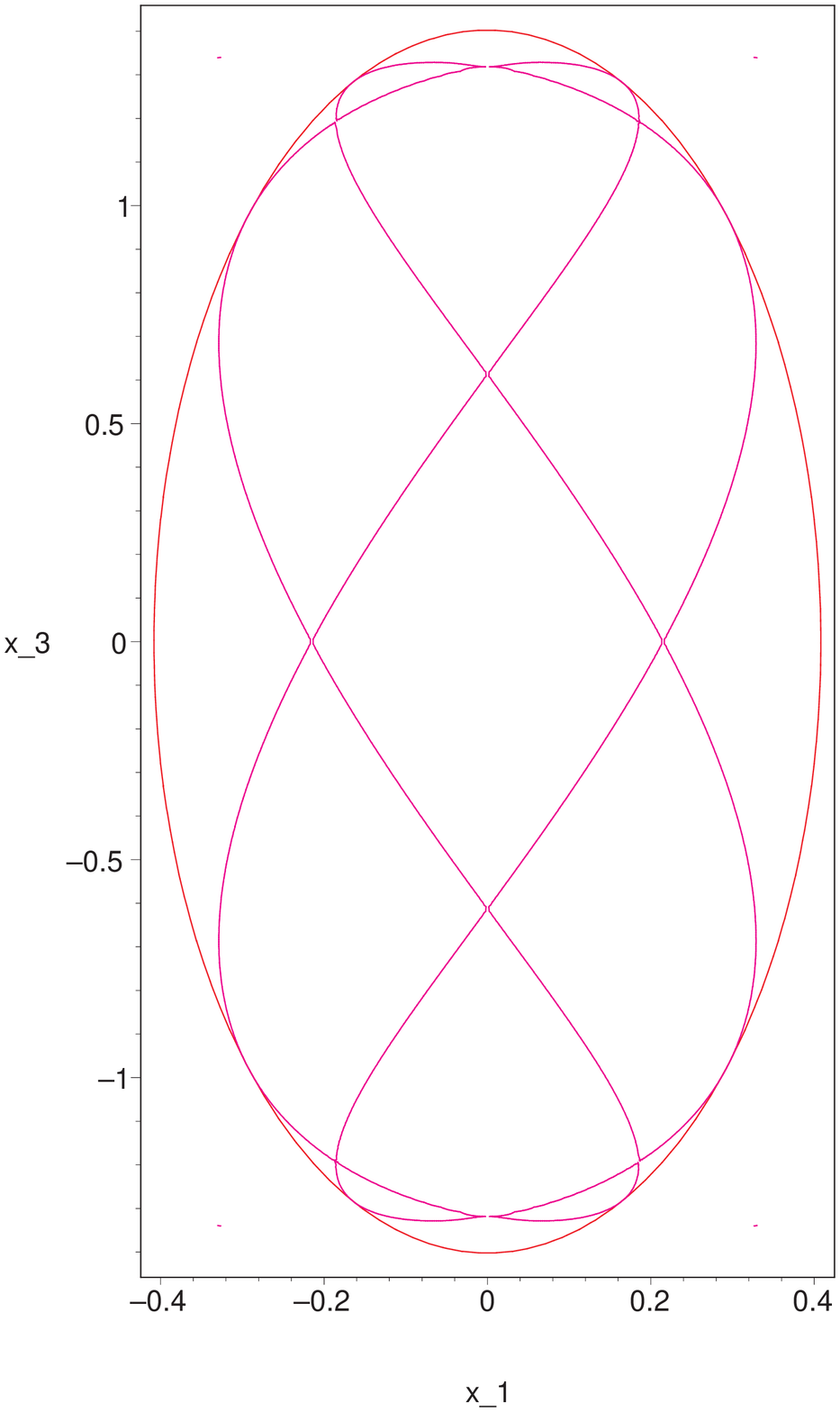}} \;
{\includegraphics[height=0.5\textwidth]{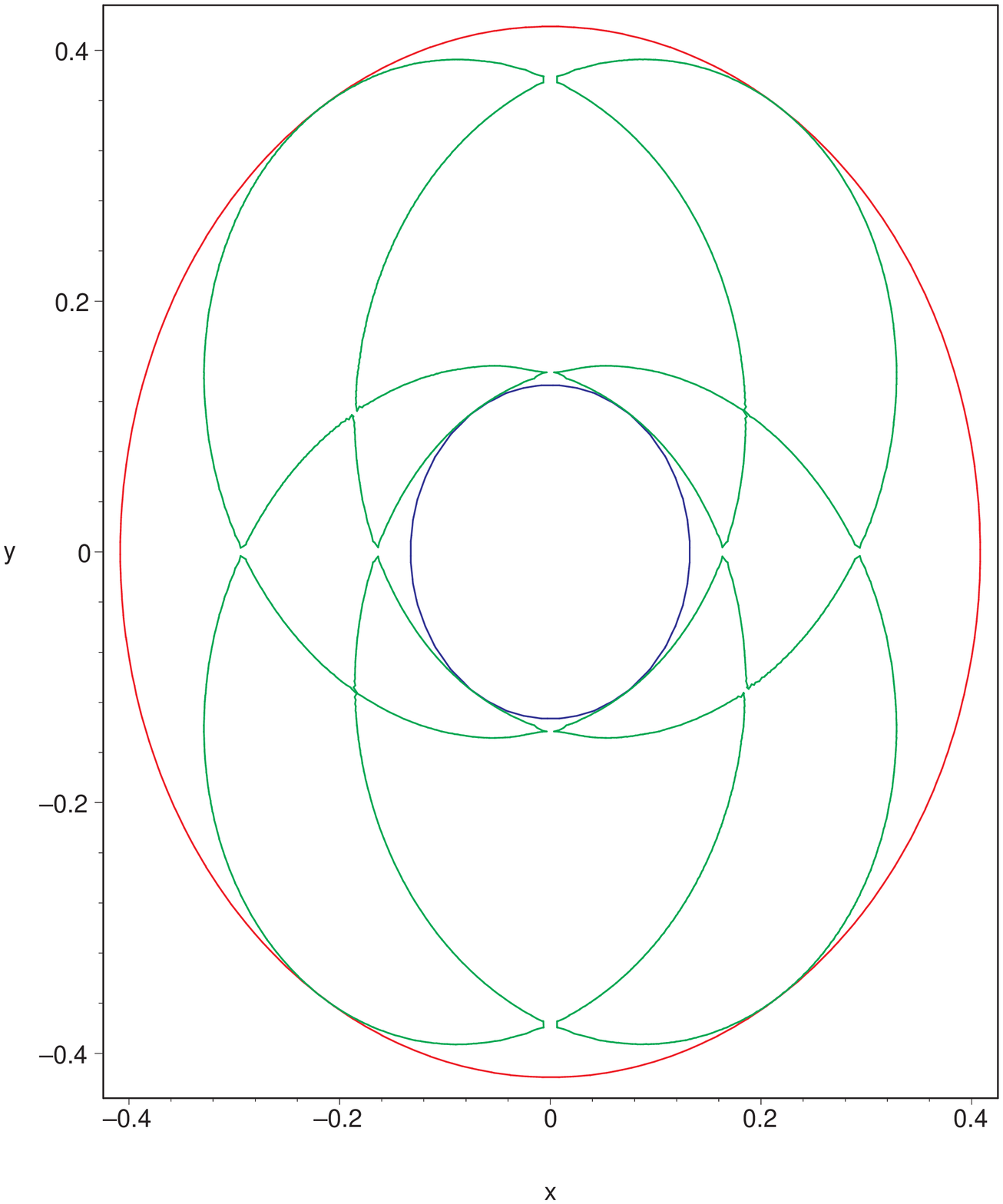}}
\end{center}
\caption{Projection of four symmetrically related closed geodesics
onto $(X_1,X_3)$ and $(X_1,X_2)$-planes for $\tilde\wp_*=-11$. In the latter case the projection
of the common caustic is indicated.}\label{gen3_1.fig}
\end{figure}

\paragraph{An Example of a Generic Closed Geodesic.} For the above numerical
choice of $g_2, g_3$ one gets $B=3$, $e_1=-0.83054$, $e_2=-0.067069$, $e_3=0.89761$ and
the formulas (\ref{choice}) (or the images of the values in (\ref{ex1})) yield
$$
a_1= 0.16667, \quad a_2 = 0.1776, \quad c= 0.3579, \quad  a_3= 1.96703.
$$
(This means that the corresponding ellipsoid is almost
"prolate"\footnote{ In all our numeric examples some of the branch
points of the curve $\Gamma$ are rather close to each other.
Apparently, this phenomenon is unavoidable and is due to the
projective transformation (\ref{LAZ}).}.) Projections of the
intersection $Q\cap {\mathcal V}_\wp$ onto $(X_1,X_3)$- and
$(X_1,X_2)$-planes
for $\tilde\wp_*=-11$ are given in Figure \ref{gen3_1.fig}.

One can see that this intersection actually consists of
{\it four} closed geodesics obtained from each other by reflections
$(X_1,X_2) \mapsto (\pm X_1,\pm X_2)$.
Each geodesics has the only self-intersection point at $X_3=0$ and
corresponds to the polhode in Figure \ref{polodias3-1.fig} (b) which is passed
{\it two times}.

It is natural to conjecture that the four geodesics are real parts
of one and the same spatial elliptic curve which are obtained from
each other via translations by elements of a finite order subgroup
of the curve.

\paragraph{Special Geodesic for ${\mathcal H}_\infty$.} In the special case $\tilde\wp_*=\infty$
the equation of the surface ${\mathcal V}_\wp$ simplifies drastically and admits the following
factorization
\begin{equation} \label{factor}
\left( \alpha ( X_{1}-\gamma)^{2}+\beta X_{3}^{2}-\delta\right)\cdot
\left( \alpha \left( X_{1}+\gamma \right)^{2}+\beta X_{3}^{2}-\delta \right )=0 ,
\end{equation}
where
\begin{equation} \label{coeff1}
\begin{aligned}
\alpha & =\frac{\sqrt{2}}{9}\left( B-6e_{1}\right)
( \left(2B+3e_{1}\right) -\sqrt{3} R ) , \\
\beta & =\frac{1}{3\sqrt{6}}\left( B+3e_{1}\right) \;
\left(R -3\sqrt{3}e_{1}\right) , \\
\gamma & =-\frac{2}{9\alpha }\left( R+3\sqrt{3}e_{1}\right) R \sqrt{B}, \\
\delta &=- \frac{B^2}{3\sqrt{2}} \frac{\left( R +3\sqrt{3}e_{1}\right)^{2}}
{\left(\sqrt{3} R-\left( 3e_{1}+2B\right) \right) \left( B-6e_{1}\right) } \, .
\end{aligned}
\end{equation}
and, as above, $B=\sqrt{3g_2}$, $R=\sqrt{3g_2-9 e_1^2}$.
\medskip

Equation (\ref{factor}) defines a union of two elliptic cylinders in ${\mathbb R}^3$
that are transformed to each other by mirror symmetry with respect to the plane $X_1=0$.
It appears that each cylinder is tangent to the ellipsoid $Q$
at a point $(X_2=X_3=0)$ and
cuts out a closed geodesic with the only self-intersection at this point. As a result,
the special closed geodesic on $Q$ related to the polhode (\ref{quadr})
is defined by its intersection with just a {\it quadratic} surface defined by one of the
two factors in (\ref{factor})).

\paragraph{Remark.}
Note that due to the self-intersection, the special geodesic in
${\mathbb R}^3$ (${\mathbb P}^3$) is a {\it rational} algebraic
curve and not an elliptic one, as the intersection of two generic
quadrics. It admits parameterization
$$
X_1=d+h_1 \cos (2\nu), \quad X_2=h_2 \sin (2\nu), \quad X_3= h_3 \sin \nu, \qquad
\nu\in {\mathbb R},
$$
$h_i,d$ being certain
constants\footnote{Here the parameter $\nu$ is not a linear function of time
$t$ or the rescaling parameter $s$ in (\ref{tau-1}).}.

On the other hand, in the phase space $(X,\dot X)$ the
corresponding periodic solution has no self-intersections and
represents an {\it elliptic} curve. Indeed, in view of formulas
(\ref{spheroconic2}), the latter can be regarded as a 4-fold
covering of the rational special polhode (\ref{quadr}). The
covering has simple ramifications at 8 points that are projected
to two vertices of the domain $S$ and two vertices of the
symmetric domain $S'$ obtained by reflection with respect to the
diagonal $z_1=z_2$. Then, according to the Riemann--Hurwitz
formula (see, e.g., \cite{Bel}), the covering has genus one. The
projection ${\mathbb C}^6=\{(X,\dot X)\}\mapsto {\mathbb
C}^3=\{X\}$ maps two different points of the elliptic solution to
the self-intersection point on $Q$.
\medskip

\begin{figure}[hb]
\begin{center}
{\includegraphics[height=0.65\textwidth]{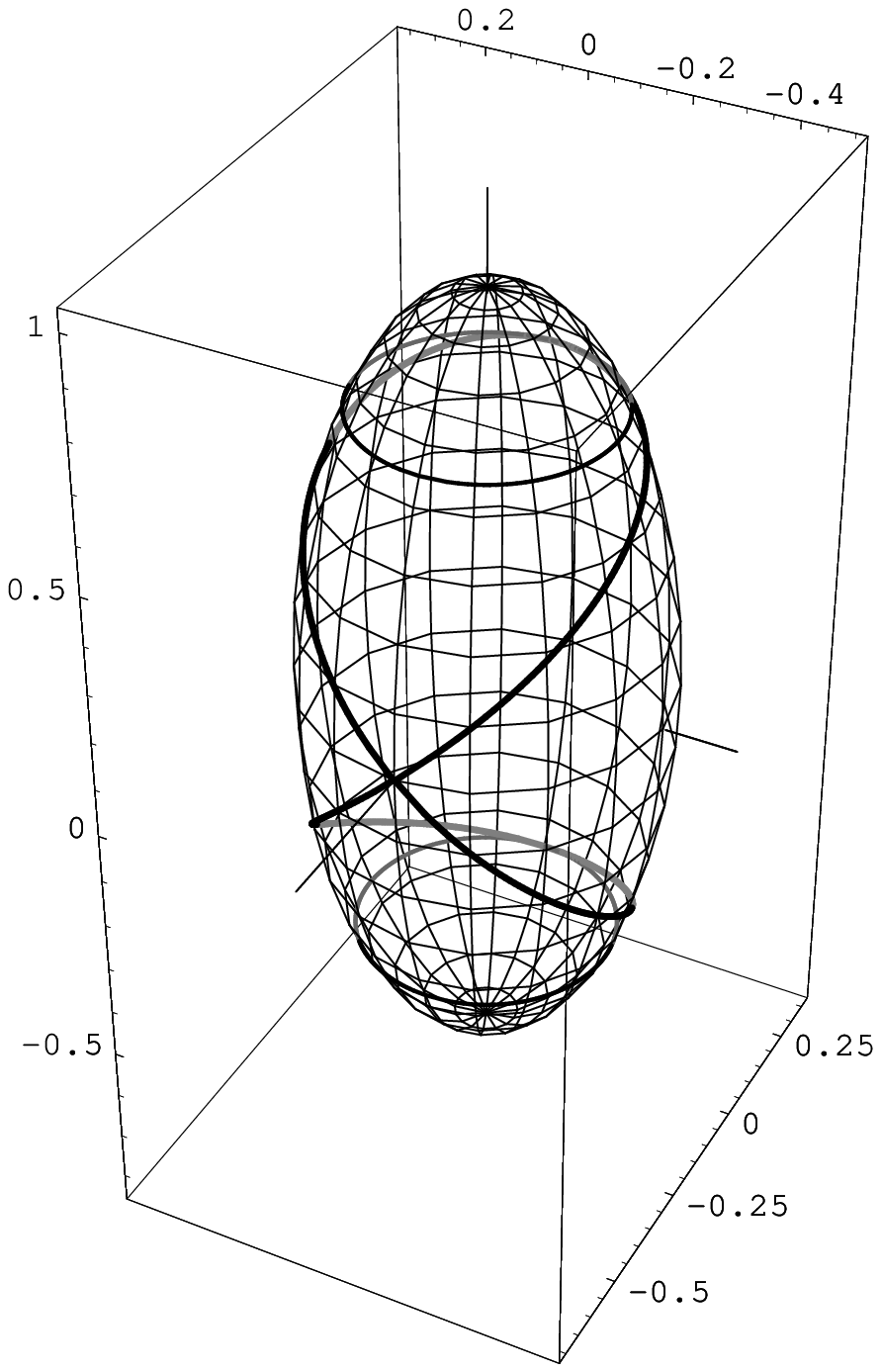}} \;
 {\includegraphics[height=0.6\textwidth]{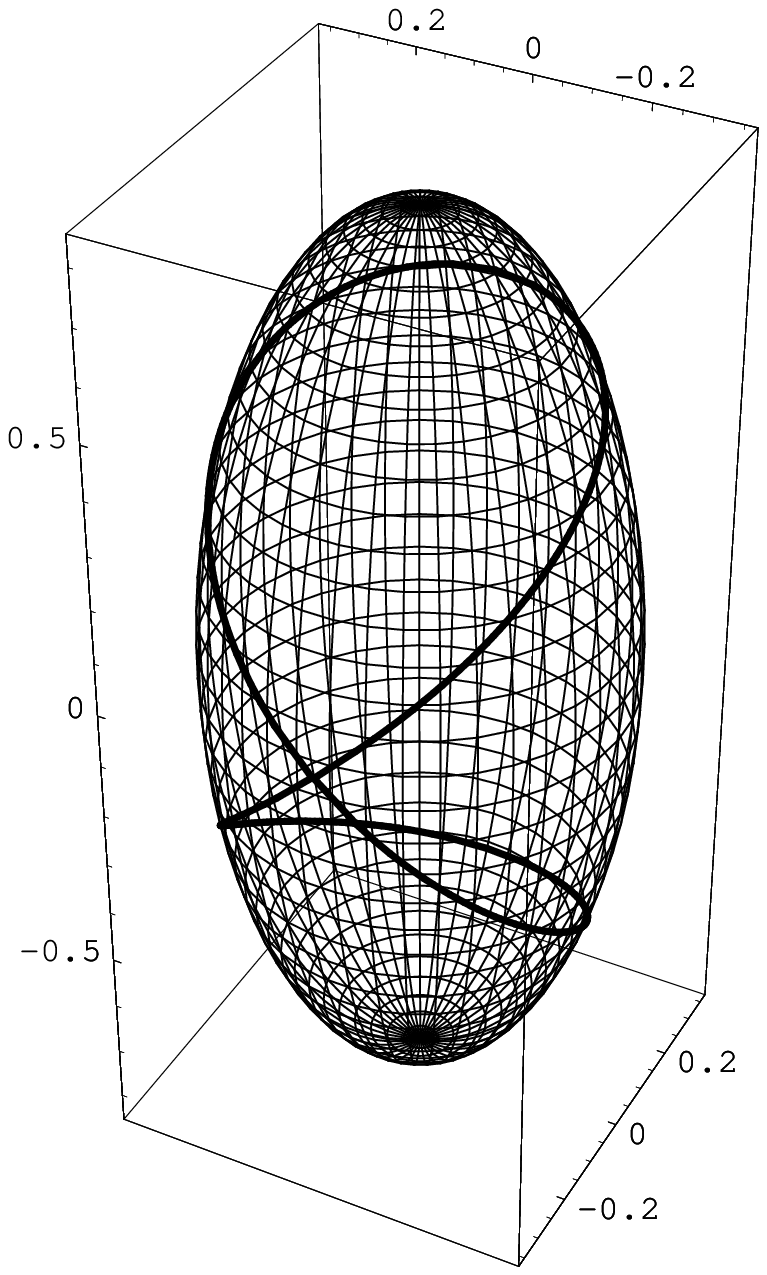}}
\end{center}
\caption{Two 3D Images of the Special Geodesic Corresponding to ${\cal H}_\infty$.}\label{sp1.fig}
\end{figure}

For the above values of the parameters $a_1, a_2,a_3, c$ the 3D graph of the special geodesic
is shown in Figure \ref{sp1.fig}.

\paragraph{Special Geodesic for ${\mathcal H}_{E_1}$.}
Applying the transformation (\ref{LAZ}) with
$b_1= -\sqrt{3g_2}$ to the special polhode (\ref{E_1}) and
making the substitution (\ref{spheroconic}) we arrive at a sextic surface
in ${\mathbb R}^3$ given by equation
\begin{align}
\quad & -256 B^{3} f_{1}^{3} X_{1}^{6}
-\left(48B^{3}f_{1}f_{2}^{2}+432Be_{1}^{2}f_{1}f_{2}^{2}
+288B^{2}e_{1}f_{1}f_{2}^{2}\right) X_{1}^{2}X_{3}^{4} \nonumber \\
& +\left( 2304B^{3}e_{1}f_{1}^{2}-192B^{4}f_{1}^{2}-6912B^{2}e_{1}^{2}f_{1}^{2}\right) X_{1}^{4}
\nonumber \\
& + \left( 192B^{4}f_{1}f_{2}+5184Be_{1}^{3}f_{1}f_{2}-864B^{3}e_{1}f_{1}f_{2}-2592B^{2}e_{1}^{2}f_{1}f_{2}\right)
 X_{1}^{2}X_{3}^{2} \nonumber \\
& +\left( -4B^{3}f_{2}^{3}-108e_{1}^{3}f_{2}^{3}-108Be_{1}^{2}f_{2}^{3}-36B^{2}e_{1}f_{2}^{3}\right) X_{3}^{6}
\nonumber \\
& +\left( 3B^{5}f_{2}-11\,664e_{1}^{5}f_{2}-27B^{4}e_{1}f_{2}+1296B^{2}e_{1}^{3}f_{2}
-108B^{3}e_{1}^{2}f_{2}\right) X_{3}^{2} \nonumber \\
&+ \left( 864B^{4}e_{1}f_{1}-46\,656Be_{1}^{4}f_{1}-36B^{5}f_{1}+31\,104B^{2}e_{1}^{3}f_{1}
-7776B^{3}e_{1}^{2}f_{1}\right) X_{1}^{2} \nonumber \\
& +\left( 1944e_{1}^{4}f_{2}^{2}-12B^{4}f_{2}^{2}+648Be_{1}^{3}f_{2}^{2}
-144B^{3}e_{1}f_{2}^{2}-324B^{2}e_{1}^{2}f_{2}^{2}\right) X_{3}^{4} \nonumber  \\
& -\left( 192B^{3}f_{1}^{2}f_{2}+576B^{2}e_{1}f_{1}^{2}f_{2}\right) X_{1}^{4}X_{3}^{2}=0,
\label{sextic} \end{align}
where
$$
f_{1}=\left( 9e_{1}^{2}-\frac{7}{4}B^{2}+B R \sqrt{3}\right), \quad
f_{2}=\left( 27e_{1}^{2}-\frac{3}{2}B^{2}-9Be_{1}+B R \sqrt{3}+3R e_{1}\sqrt{3}\right).
$$
It cuts out a pair of closed geodesic on $Q$ that are transformed to each other by mirror
symmetry with respect to the plane $X_2=0$. Both geodesics have a 3D shape similar to that
in Figure \ref{sp1.fig}, each of them has the only self-intersection point for $(X_1=X_3=0)$.

Note, however, that in contrast to quartic equation (\ref{factor}), the sextic polynomial
in (\ref{sextic}) does not admit a factorization, hence none of the above geodesics can be
represented as the intersection of $Q$ with a quadratic or a cubic cylinder.

\begin{figure}[hb]
\begin{center}
\includegraphics[height=0.55\textwidth,width=0.4\textwidth]{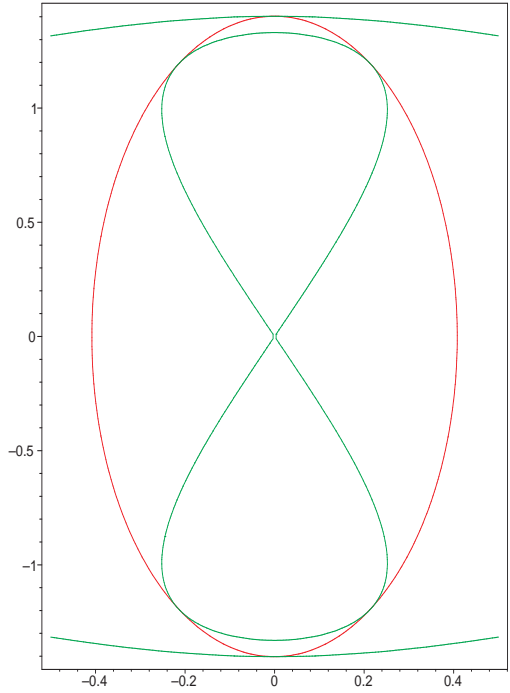}
\end{center}
\caption{Projection of the sextic surface (\ref{sextic}) and the geodesic for ${\cal H}_{E_1}$ onto
$(X_1,X_3)$-plane.}\label{sp2.fig}
\end{figure}

For the above choice of moduli $g_2, g_3$ and the parameters $a_i, c$ the projection
of the sextic surface and the corresponding geodesics onto $(X_1,X_3)$-plane
are given in Figure \ref{sp2.fig}.

\newpage

\paragraph{Remark.} As follows from the above considerations, under the condition
$a_1< a_2<c<a_3$ all the real closed geodesics of the one-parametric family have one self-intersection
point on the equator $\{X_3=0\}\subset Q$,
and as the parameter $\tilde\wp_*$ ranges from $-\infty$ to $E_1$,
this point varies from the $X_1$-axis to $X_2$-axis.
\medskip

The case $a_1< c< a_2 <a_3$ will be illustrated in detail elsewhere.

\section{The case of 4:1 tangential covering}
This case was originally studied by Darboux 
and later appeared in
paper \cite{TV} 
in connection with new elliptic solutions of the KdV equation (see also \cite{Tr,Smirnov2}).
Namely, the genus 2 curve $G$
\begin{equation} \label{4:1}
w^2 = -(z-6e_1) \prod_{l=1}^4 (z-z_l) ,
\end{equation}
with
\begin{align*} 
z_{1,2}  & = e_3 +2e_2 \pm \sqrt{28e_2^2+76e_2e_3+40e_3^2}, \\
z_{3,4}  & = e_2 +2e_3\pm\sqrt{28e_3^2+76e_2e_3+40e_2^2},
\end{align*}
is a 4-fold cover of the curve (\ref{ell1}).
It also covers second elliptic curve \\
${\cal E}_{2} =\left\{ W^{2}= -4 (Z-E_1) (Z -E_2) (Z -E_3) \right\}$,
such that
\begin{align}
E_{1} & =3\left( e_{2}-e_{3}\right) ^{3}-6\left( e_{2}-e_{1}\right) (5e_{1}-2e_{3})^{2},
\nonumber \\
E_{2} &=-6\left( e_{2}-e_{3}\right)^{3}+3\left( e_{2}-e_{1}\right)(5e_{1}-2e_{3})^{2},
\label{ES} \\
E_{3} &=3\left( e_{2}-e_{3}\right) ^{3}+3\left(
e_{2}-e_{1}\right)(5e_{1}-2e_{3})^{2} \nonumber
\end{align}
as described by one of the formulas
\begin{align}\label{deg4}
Z & = E_{1}+ F_\alpha(z), \quad F_\alpha=\frac{9}{4}\frac{\left( z^{2}-3e_{\alpha}z-24\left(
e_{\beta}^{2}+e_{\gamma}^{2}\right) -51 e_{\beta}e_{\gamma}\right) ^{2}}{z-6e_{\alpha}} , \\
W &= - w \frac {d Z(z)}{dz} , \nonumber
\end{align}
where $(\alpha,\beta,\gamma)$ is a circular permutation of $(1,2,3)$.
In the sequel we assume \\ $\alpha=1,\beta=2,\gamma=3$.

Substituting projection formulas (\ref{deg4}) to the generating
equation (\ref{main}) one obtains equation of generic polhodes of
degree 16, which is much more tedious than the family
(\ref{final-3}) for the 3:1 cover, so we do not give it here.

\paragraph{The special polhodes for $\tilde\wp_*=\infty$ and $\tilde\wp_*=E_1$.}
Substituting projection formulas (\ref{deg4}) to the special generating
equation (\ref{infty}), we obtain algebraic equation of degree 4,
\begin{align}
& 6e_{1}\left(z_{1}^{3}+z_{2}^{3}\right) -36e_{1}^{2}\left( z_{1}^{2}+z_{2}^{2}\right)
-z_{1}^{2}z_{2}^{2}+ \left(12e_{1} (z_{1}+z_{2}) -z_{1}^{2}-z_{2}^{2}\right)z_{2}z_{1}
\nonumber \\
 & + (6 e_{2}e_{3}+3 e_{1}^{2})z_{1}z_{2}
 + \left(54e_{1}^{3}-612e_{1}e_{2}e_{3}-288e_{1}e_{2}^{2}-288e_{1}e_{3}^{2}\right)(z_{1}+ z_{2})
 \nonumber \\
& +9\left( 17e_{2}e_{3}+8e_{2}^{2}+8e_{3}^{2}\right) \left(
17e_{2}e_{3}+12e_{1}^{2}+8e_{2}^{2}+8e_{3}^{2}\right) =0 \, . \label{sp4-infty}
\end{align}

Next, substituting (\ref{deg4}) into the special generating
equation (\ref{sp_gen}) with $E_\alpha=E_1$ and taking into account
relation $E_1+E_2+E_3=0$, one gets the following equation of degree 8
\begin{gather}
81\,z_1^{4}\,z_2^{4}  - 486e_1(z_2^{3}\,z_1^{4} + z_2^{4}\,
z_1^{3}) -({3159}\,e_{2}^{2}+6804\,{e_{2}}\,{e_{3}} + {3159}\,{e_{3}}^{2})\,(z_2^{2}
\,z_1^{4} + z_2^{4}\,z_1^{2}) \nonumber \\
+ 2916\,e_1^{2}\,z_1^{3} z_2^{3} + 2916\,e_{1} \xi_1 (z_2\,z_1^{4} + z_2^{4}\,z_1)
+ 1458\,e_1 \xi_2 \,(z_2^{2}\,z_1^{3} + z_2^{3}\,z_1^{2}) \nonumber \\
+ 729\,\xi_1^{2}\,(z_1^{4} + z_2^{4}) - 8748\,\xi_1\,e_1^{2}\,(z_2\,z_1^{3} + z_2^{3}\,z_1)
+ 729\,\xi_2^{2}\,z_1^{2}\,z_2^{2}  - 4374 e_1 \xi_1^{2}\,(z_1^{3} + z_2^{3})  \nonumber \\
- 4374\,e_1 \xi_1\xi_2 \,(z_2\,z_1^{2} + z_2^{2}\,z_1)
 - 2187\,\xi_2\,\xi_1^{2}\,(z_1^{2} + z_2^{2})
+ 16\bigg (104976\,{e_{2}}^{6} + 656100\,{e_{2}}^{5}\,{e_{3}} \nonumber \\
+ \frac {6725025}{4} \,{e_{2}}^{4}\,{e_{3}}^{2}
+ \frac {4520529}{2} \,{e_{2}}^{3}\,{e_{3}}^{3}
+ \frac {6725025}{4} \,{e_{2}}^{2}\,{e_{3}}^{4} + 656100\,{e_{2}}\,{e_{3}}^{5}
+ 104976\,{e_{3}}^{6} - 2\,{E_{2}}^{2} \nonumber \\
- 5\,{E_{2}}\,{E_{3}} - 2\,{E_{3}}^{2}\bigg )z_1\,z_2
+ \frac {3}{8} e_1 (1119744\,{e_{2}}^{6} + 7138368\,{e_{2}}^{5}\,{e_{3}} +
18528264\,{e_{2}}^{4}\,{e_{3}}^{2} \nonumber \\
 + 25021467\,{e_{2}}^{3}\,{e_{3}}^{3} + 18528264\,{e_{2}}^{2}\,{e_{3}}^{4} + 7138368\,{e_{2}}\,{e_{3}}^{5}
 + 1119744\,{e_{3}}^{6} + 32\,{E_{2}}^{2} \nonumber \\
+ 80\,{E_{2}}\,{E_{3}} + 32\,{E_{3}}^{2})(z_1+z_2)
+ 52225560\,e_{2}^{6}\,e_{3}^{2} + 107298594\,e_{2}^{5}e_{3}^{3}
+ \frac {2165451489}{16} \,e_{2}^{4}e_{3}^{4} \nonumber \\
 + 14276736\,{e_{2}}^{7}\,{e_{3}} - 144\,{e_{2}}\,{e_{3}}\,{E_{3}}^{2} - 72\,{e_{3}}^{2}\,{E_{2}}^{2}
- 180\,{e_{3}}^{2}\,{E_{2}}\,{E_{3}} - 72\,{e_{3}}^{2}\,{E_{3}}^{2} \nonumber \\
+ 107298594\,{e_{2}}^{3}\,{e_{3}}^{5} + 52225560\,{e_{2}}^{2}\,{e_{3}}^{6}
 + 14276736\,e_{2}e_{3}^{7} + 1679616\,{e_{3}}^{8}
 - 72\,{e_{2}}^{2}\,{E_{2}}^{2} \nonumber \\
-180\,{e_{2}}^{2}\,{E_{2}}\,{E_{3}} - 72\,{e_{2}}^{2}\,{E_{3}}^{2}
- 144\,{e_{2}}\,{e_{3}}\,{E_{2}}^{2} - 360\,{e_{2}}\,{e_{3}}\,{E_{2}}\,{E_{3}}
 + 1679616\,{e_{2}}^{8} =0,
\label{sp4_E_1} \\
\xi_1= 8\,{e_{2}}^{2} + 17\,{e_{2}}\,{e_{3}} + 8\,{e_{3}}^{2}, \quad
\xi_2= 13\,{e_{2}}^{2} + 28\,{e_{2}}\,{e_{3}} + 13\,{e_{3}}^{2} ,\nonumber
\end{gather}

\paragraph{Examples of polhodes in $S$.} To illustrate the above polhodes,
in the first elliptic curve (\ref{ell1}) we choose
$$
e_1=-2, \quad e_2= -1, \quad e_3=3, \mbox{  so that   }\quad
E_1=-1728, \quad E_{2}= 1152, \quad E_{3}=576 ,
$$
and the roots of the polynomial in (\ref{4:1}) become
\begin{equation} \label{roots4}
(-12.0, \; -11.649, \; -3.0, \; 13.0, \; 13.649).
\end{equation}
As a result, the square domain $S\subset(z_1,z_2)$
where the corresponding variables $w_1, w_2$ are real is
\begin{equation}\label{real-4}
S= \{ -11.649 \le z_1 \le -3.0, \quad 13.0 \le z_2 \le 13.649 \}.
\end{equation}

If the parameter $ \tilde\wp_*$ ranges in the interval $(-\infty; E_1= -1728)$, then
the conditions (\ref{roots}) do not hold in $S$, hence the real
asymmetric part of the polhode is non-empty.

\begin{figure}[hb]
\begin{center}
\subfigure[$\tilde\wp_* =-4900$]{
\includegraphics[height=0.25\textwidth]{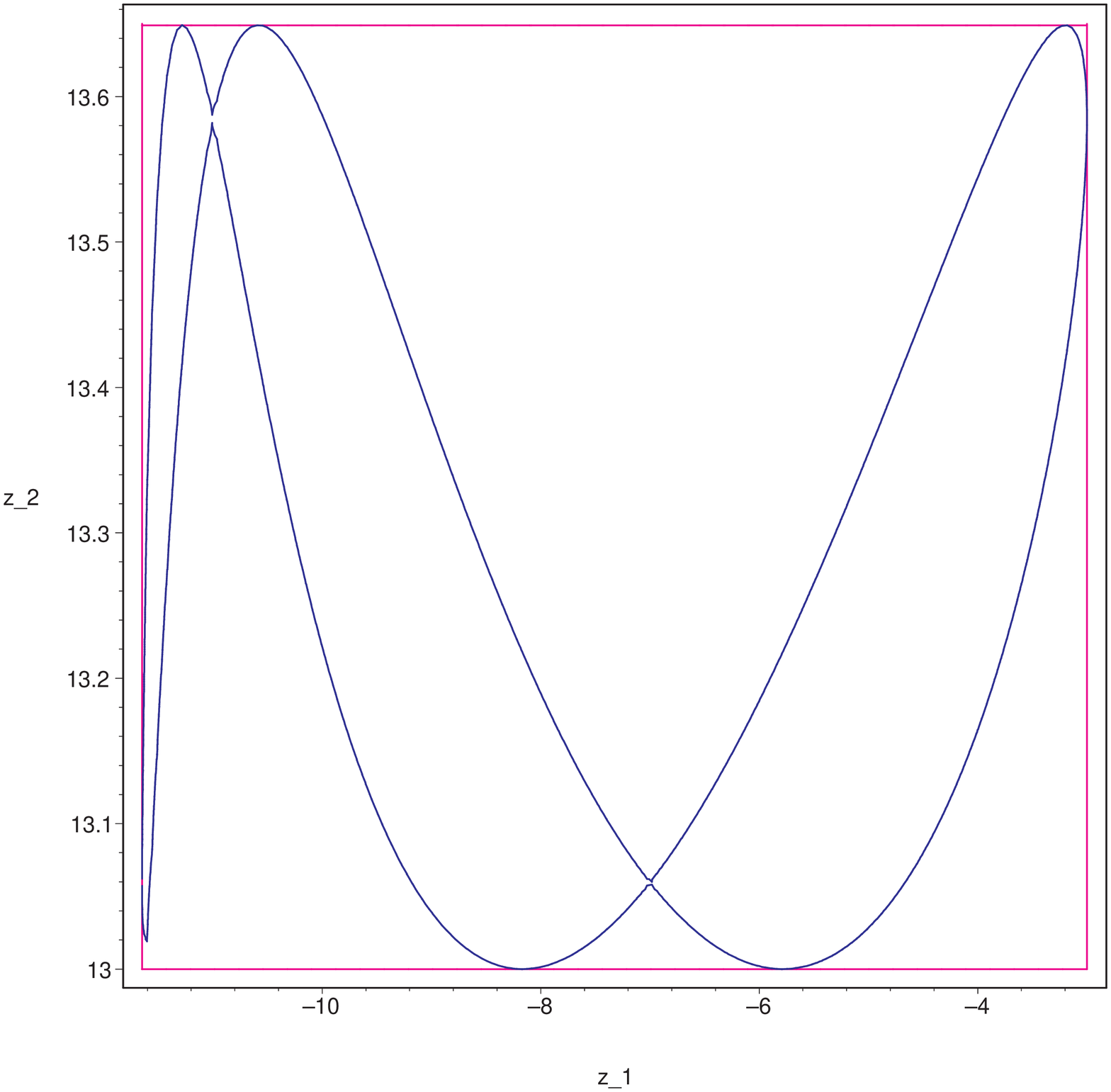}} \quad
\subfigure[$\tilde\wp_*=-7000$]{
\includegraphics[height=0.25\textwidth]{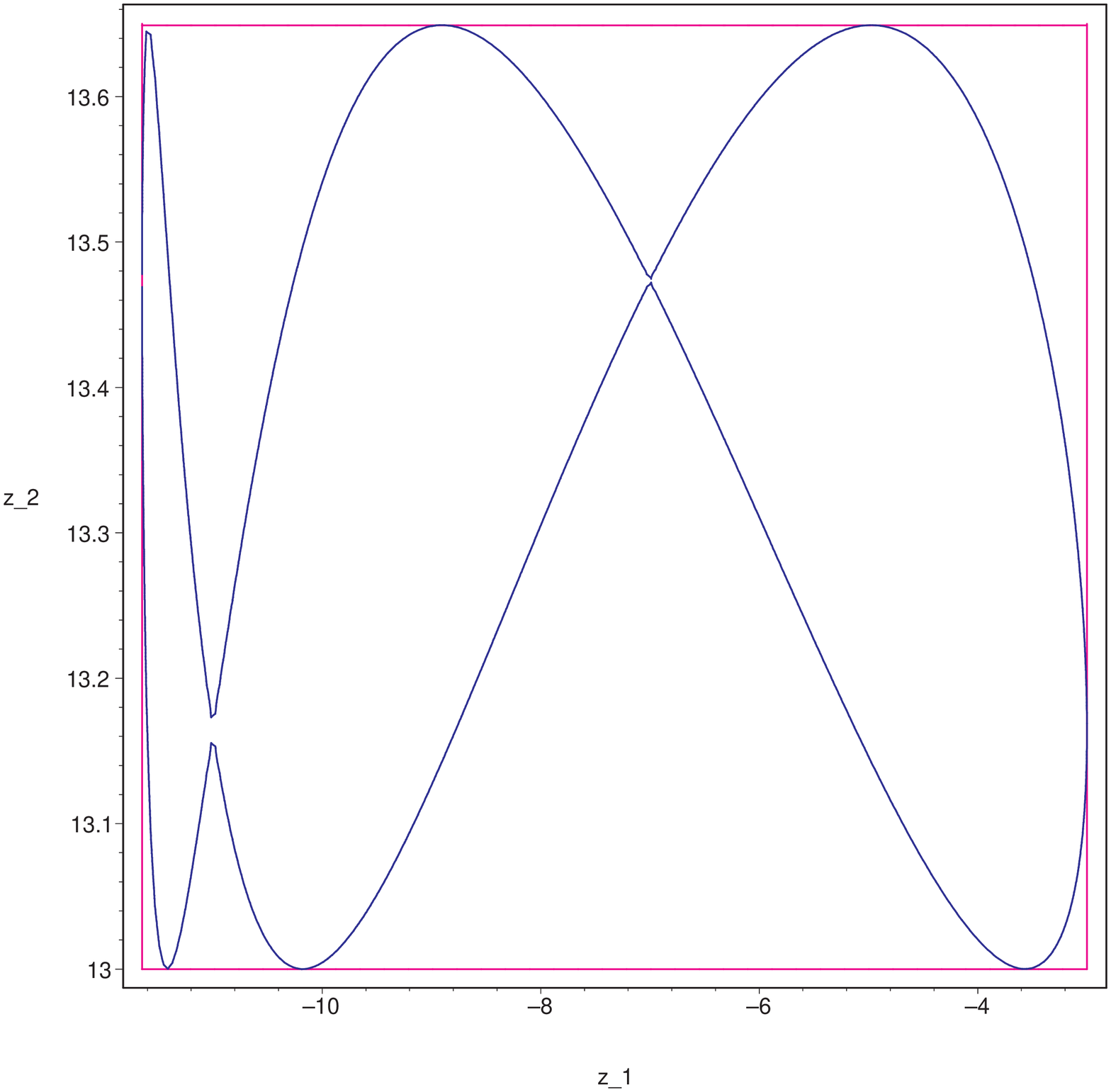}}
\end{center}
\caption{Generic polhodes in $S\subset {\mathbb R}^2=(z_1, z_2)$}
\label{polodias4-1.fig}
\begin{center}
\subfigure[$\tilde\wp_* =-\infty$]{
\includegraphics[height=0.25\textwidth]{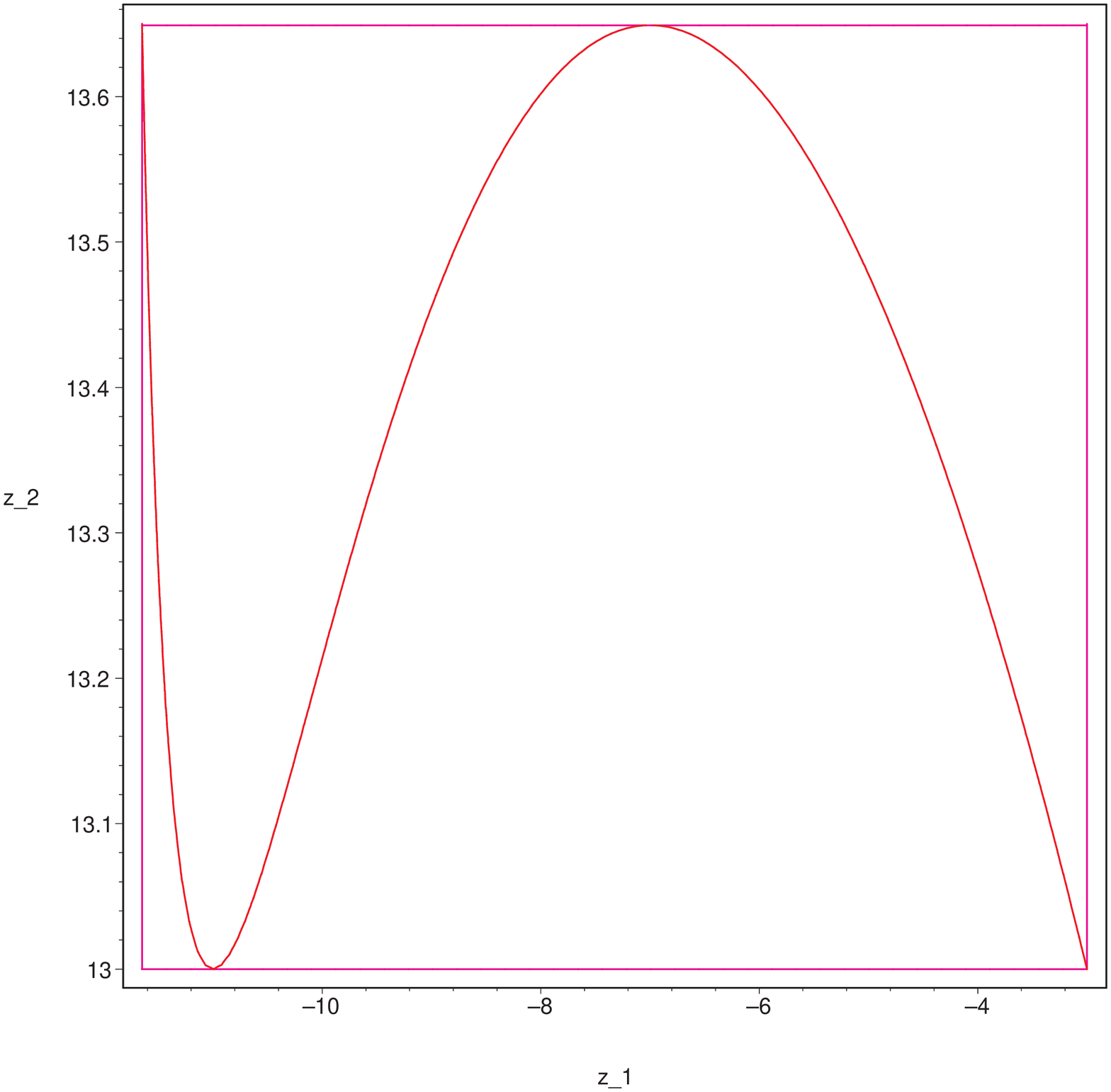}} \quad
\subfigure[$\tilde\wp_* =-1728$]{
\includegraphics[height=0.25\textwidth]{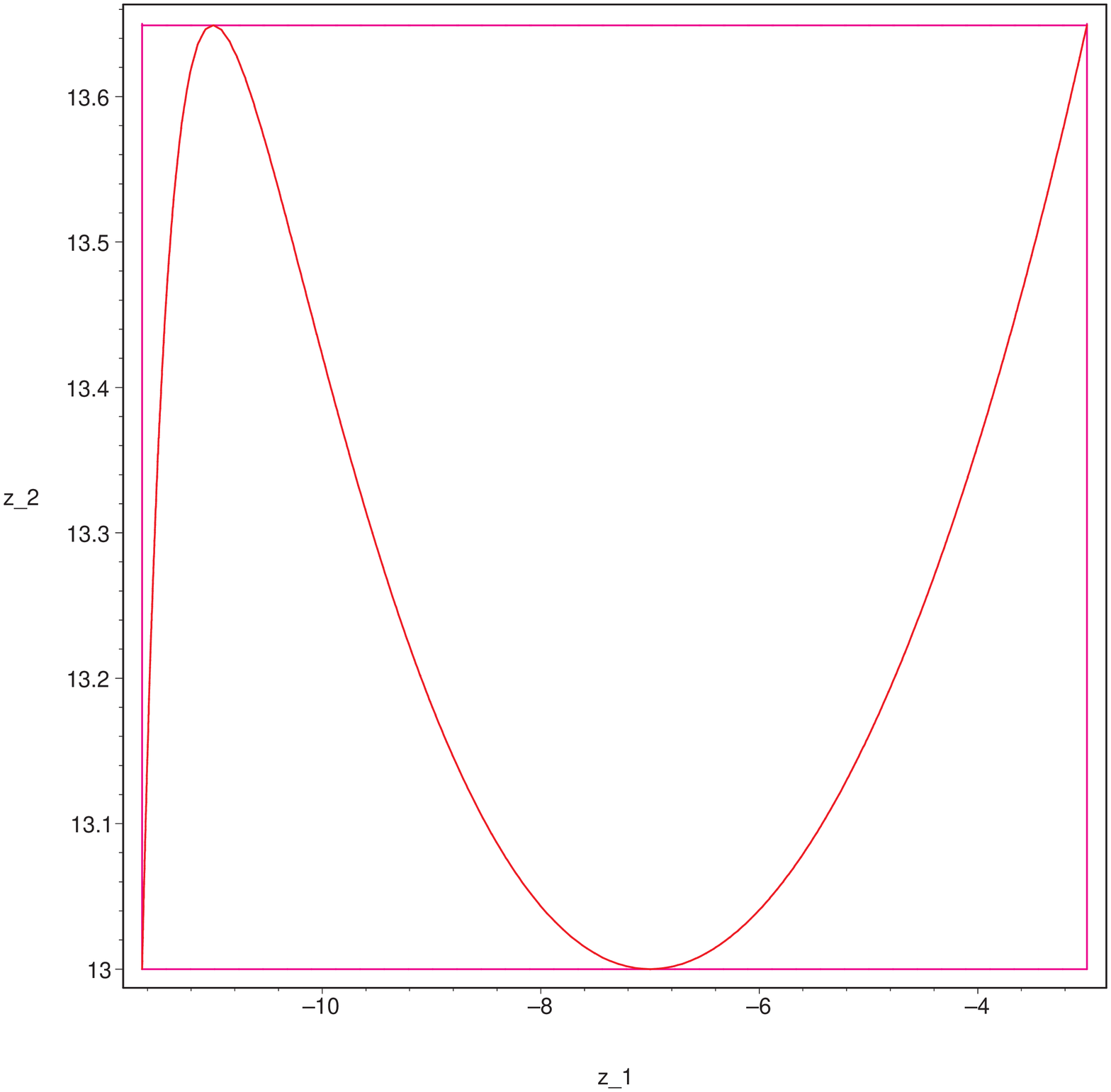}}
\end{center}
\caption{Special polhodes.}\label{limit.pol4.fig}
\end{figure}

Graphs of polhodes 
in the domain (\ref{real-4}) for a generic
value of $\tilde\wp_*\in (-\infty; -1728)$ is given in Figure \ref{polodias4-1.fig},
whereas the special polhodes defined by (\ref{sp4-infty}) and (\ref{sp4_E_1})
are shown in Figure \ref{limit.pol4.fig}.

As seen from Figure \ref{polodias4-1.fig}, the generic polhodes in $S$
intersect generic lines $z_2=$const and $z_1=$const at 6 and 2 points respectively.

\newpage


\paragraph{Special Closed Geodesic for ${\cal H}_\infty$.} Assuming, as above,
$e_1 < e_2< e_3$, we conclude that $b_1=- 6 e_1$ is the minimal root of the hyperelliptic
polynomial in (\ref{4:1}). Setting this value into the transformation (\ref{LAZ}), we get
$$
z_{1}=\frac{6e_{1}\lambda _{1}+1}{\lambda _{1}}, \quad
z_{2}=\frac{6e_{1}\lambda _{2}+1}{\lambda _{2}}.
$$
The latter substitution transforms equation (\ref{sp4-infty}) of polhode ${\cal H}_\infty$ to
\begin{align}
 & \left( 102e_{2}e_{3}-117e_{1}^{2}+48e_{2}^{2}+48e_{3}^{2}\right)
\lambda_{1}^{2}\lambda_{2}^{2}-\lambda_{1}^{2}-\lambda _{2}^{2}-\lambda _{1}\lambda _{2} \nonumber \\
& \quad -18e_{1}\left( \lambda _{1}+\lambda _{2}\right)\lambda_{2}\lambda_{1}
 + 9\left(17e_{2}e_{3}-6e_{1}^{2}+8e_{2}^{2}+8e_{3}^{2}\right)^{2}
\, \lambda_{1}^{3}\lambda_{2}^{3}=0,
\end{align}
which describes the closed geodesic in ellipsoidal coordinates on $Q$.
Next, assuming that $a_1< a_2<c<a_3$, we find
\begin{gather*}
a_1 = \frac {1}{6 e_1+ e_2 +2e_3+\sqrt{28e_3^2+76e_2e_3+40e_2^2} }, \quad
a_2 = \frac {1}{6 e_1+ e_2 +2e_3-\sqrt{28e_3^2+76e_2e_3+40e_2^2}}, \\
a_3 = \frac {1}{6 e_1+ e_3 +2e_2 -\sqrt{28e_2^2+76e_2e_3+40e_3^2}}, \quad
c = \frac {1}  {6 e_1+ e_3 +2e_2 +\sqrt{28e_2^2+76e_2e_3+40e_3^2} }.
\end{gather*}
Applying formulas (\ref{spheroconic}) and simplifying, one obtains
equation of cylinder surface ${\cal V}_\infty$ of degree 6 in
coordinates $(X_1,X_3)$, which admits the factorization
\begin{gather}
F_{-}(X_1,X_3)\, F_{+}(X_1,X_3)=0, \nonumber \\
F_{\pm} = \pm h_{30} X_1^3 + h_{03} X_3^3 + h_{21} X_1^2 X_3
\pm h_{12} X_1 X_3^2 \pm h_{10} X_1+h_{01} X_3 , \label{cubic1}
\end{gather}
where $h_{ij}$ are rather complicated expressions of $e_2,e_3$, so we do not give
them here.

Thus ${\cal V}_\infty$ is a union of two cubic cylinders
${\cal V}_{\infty-}$, ${\cal V}_{\infty+}$ that are obtained from each other
by mirror symmetry with respect to the plane $X_1=0$.

\begin{figure}[float]
\begin{center}
{\includegraphics[height=0.55\textwidth, width=0.3\textwidth]{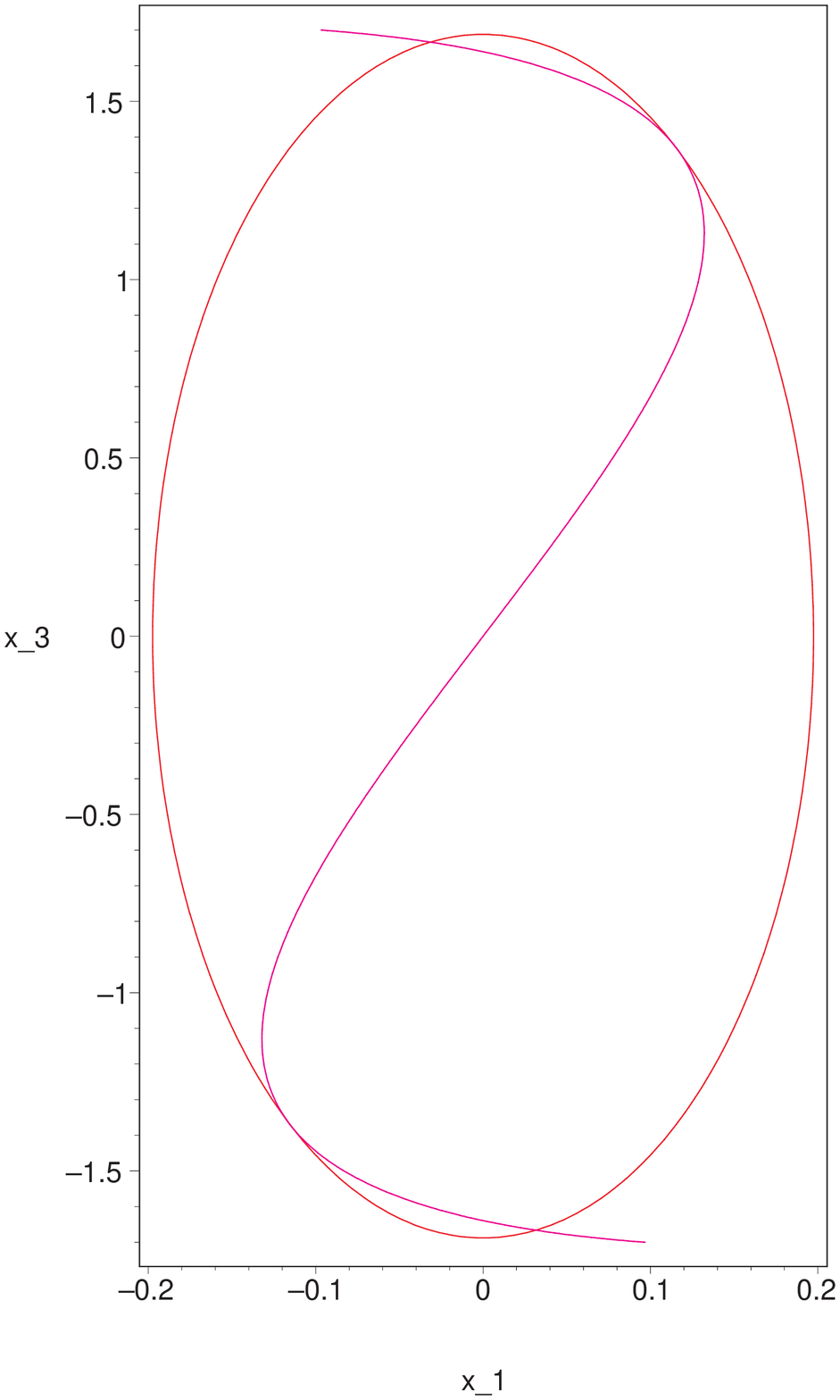}, \;
\includegraphics[height=0.8\textwidth, width=0.55\textwidth]{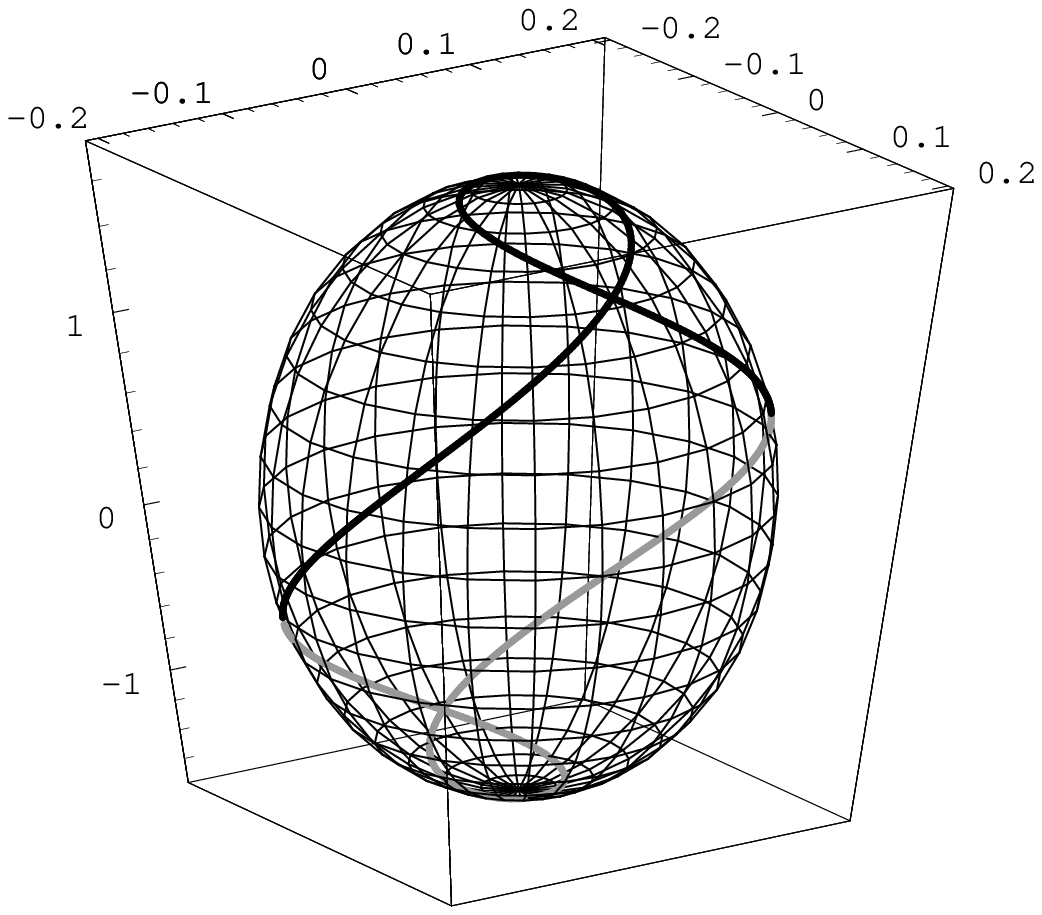} }
\end{center}
\caption{Projection of the cylinder ${\cal V}_{\infty+}$ onto $(X_1,X_3)$-plane and
Special Geodesic for ${\cal H}_\infty$.}\label{special4_1.fig}
\end{figure}

\paragraph{Example of a closed geodesic for ${\cal H}_\infty$.}
Under the above assumption $a_1< a_2<c<a_3$, the values (\ref{roots4}) lead to numbers
$$
a_{1}=3.69877\times 10^{-2}, \quad a_{2}=0.04, \quad c=0.111111, \quad
a_{3}=2.84981
$$
and the equation of the cylinder ${\cal V}_{\infty+}$ becomes
\begin{equation} \label{num3}
  -0.301917 X_{1}^{3}-0.470556 X_{1}X_{3}^{2}+0.652885 X_{1}^{2}X_{3}
+1.87228 X_{1} -0.303831 X_{3}+0.113051 X_{3}^{3} =0 .
\end{equation}
Its projection 
onto $(X_1, X_3)$-plane
and the 3D graph of the corresponding geodesic are shown in Figure \ref{special4_1.fig}
\footnote{The graph was actually produced by direct numeric integration of the
corresponding geodesic equation with initial conditions prescribed by (\ref{num3})}.
The cylinder is tangent to the ellipsoid $Q$ at 2 points with $X_2=0$, which implies that
the geodesic has two self-intersection points.

\paragraph{Remark.} The special geodesic can be regarded as a two-fold covering of
the plane algebraic curve ${\mathcal F}_\pm=\{F_{\pm}(X_1,X_3)=0\}$,
which is ramified at two points of {\it transversal} intersection
of ${\mathcal F}_\pm$ with the ellipse $\{X_1^2/a_1+X_3^2/a_3=1\}$.
(There is no ramification at the two points of tangency with the ellipse.)
Using explicit expressions of (\ref{cubic1}),
one can show that for any value of $e_2,e_3$ the genus
of ${\mathcal F}_\pm$ equals zero.
Hence, according to the Riemann--Hurvitz formula,
the special closed geodesic is a rational curve. 

However, similarly to the special geodesic for the 3:1 covering,
in the phase space $(X,\dot X)$ the periodic solution
corresponding to ${\cal H}_\infty$ has no self-intersections and
represents a connected component of real part of an elliptic
curve.

\paragraph{General closed geodesics.} For a generic parameter $\tilde\wp_*$ the equation
of the cutting cylinder ${\cal V}_{\tilde\wp_*}$ has degree 16 and
its projection onto $(X_1,X_3)$-plane looks quite tangled: it
includes 2 intersecting closed geodesics obtained from each other
by reflections $X_1 \mapsto -X_1$. The structure of the real
asymmetric part of generic and the special polhodes implies the
following behavior under the condition $a_1< a_2<c<a_3$:  all the
closed geodesics of the family have two centrally symmetric
self-intersection points and as the parameter $\tilde\wp_*$ ranges
from $-\infty$ to $E_1$, these points vary from the plane
$\{X_2=0\}$ to $\{X_1=0\}$.  All the geodesics have rotation
number $\rho=3:1$.

Similarly to the case of the 3-fold tangential covering, one can consider the ordering \\
$a_1<c< a_2<a_3$, which leads to generic and special closed geodesics on $Q$ without
self-intersections.
\newpage

\subsection*{Conclusion} In this paper we proposed a simple method of explicit
constructing families of algebraic closed geodesics on triaxial
ellipsoids, which is based on properties of tangential coverings
of an elliptic curve and the addition theorem for elliptic
functions. We applied the method to the cases of 3- and 4-fold
coverings and gave concrete examples of algebraic surfaces that
cut such closed geodesics. The latter coincide with those obtained
by direct numeric integration of the geodesic equation. This
serves as an ultimate proof of correctness of the method.
Depending on how one chooses the caustic parameter $c$ in the
interval $(a_1,a_3)$, the closed geodesics may or may not have
self-intersections.

Thus, our approach can be regarded as a useful application of the
Weierstrass--Poincar\'e reduction
theory. 
One must only know
explicit covering formulas (\ref{covs}), as well as expressions for $z$-coordinates
of the finite branch points of the genus 2 curve $G$ in terms of the moduli $g_2, g_3$.
To our knowledge, until now such expressions are calculated only for $N\le 8$.

Since the method essentially uses the algebraic addition law on the second elliptic curve ${\mathcal E}_2$,
it does not admit a straightforward generalization to similar description of 
algebraic closed geodesics on $n$-dimensional ellipsoids: as mentioned in Section 3, in this case
${\mathcal E}_2$ is replaced by an Abelian subvariety ${\mathcal A}_{n-1}$, for which 
an algebraic description is not known.

On the other hand, one should not exclude the existence of algebraic closed geodesics on
ellipsoids related to other type of
doubly periodic solutions of the KdV equation, e.g., elliptic not $x$, but in $t$-variable.
This is expected to be a subject of a separate study.

Our approach can equally be applied to describe elliptic solutions
of other integrable systems linearized on two-dimensional hyperelliptic
Jacobians or their coverings.

\subsection*{Acknowledgments}
I am grateful to A. Bolsinov, L. Gavrilov, V. Enolskii, E.
Previato, and A. Treibich for stimulating discussions, as well as
to A. Perelomov for some important suggestions during preparation
of the manuscript and indicating me the reference \cite{Braun}. I
also thank the referees for their valuable remarks that helped to
improve the text.


The support of grant BFM 2003-09504-C02-02 of
Spanish Ministry of Science and Technology is gratefully acknowledged.

\end{document}